\documentclass{ar2e}
\usepackage{graphicx}
\begin{document}

\input epsf.def   

\input psfig.sty

\jname{Annu. Rev. Phys. Chem}
\jyear{2007}
\jvol{39}
\ARinfo{1056-8700/97/0610-00}

\title{Fluctuation Theorems}

\markboth{Sevick, Prabhakar, Williams \& Searles}{Fluctuation Theorems}

\author{E.M. Sevick, R. Prabhakar, Stephen R. Williams \affiliation{Research School of Chemistry, Australian National University, Canberra, Australian Capital Territory 0200, Australia; email: sevick@rsc.anu.edu.au, Prabhakar@rsc.anu.edu.au, swilliams@rsc.anu.edu.au}  Debra J. Searles\affiliation{Nanoscale Science and Technology Centre, School of Biomolecular and Physical Sciences, Griffith University, Brisbane, Qld 4111, Australia; email: D.Bernhardt@griffith.edu.au} 
}


\begin{abstract}
Fluctuation theorems, which have been developed over the past 15 years, have resulted in fundamental breakthroughs in our understanding of how irreversibility emerges from reversible dynamics, and have provided new statistical mechanical relationships for free energy changes.  They describe the statistical fluctuations in time-averaged properties of many-particle systems such as fluids driven to nonequilibrium states, and provide some of the very few analytical expressions that describe nonequilibrium states.    Quantitative predictions on fluctuations in small systems that are monitored over short periods can also be made, and therefore the fluctuation theorems allow thermodynamic concepts to be extended to apply to finite systems.  For this reason, fluctuation theorems are anticipated to play an important role in the design of nanotechnological devices and in understanding biological processes.  These theorems, their physical significance and results for experimental and model systems are discussed. 
\end{abstract}

\maketitle

\section{Introduction}
Thermodynamics is the study of the flow of heat and the transformation of work into heat.  Our understanding of thermodynamics is largely confined to equilibrium states.  Linear irreversible thermodynamics is  an extension of the 19th century concepts of equilibrium thermodynamics to systems that are sufficiently 
close to equilibrium that intensive thermodynamic variables can be approximated by the same functions of local state variables, as would be the case if the entire system was in complete thermodynamic equilibrium \cite{DeGroot-Mazur, Kondepudi-Prigogine}.   Moreover these traditional concepts are limited in application to large systems or averages over an ensemble of states, referred to as the ``thermodynamic limit".

Inventors and engineers endeavour to scale down machines and devices to nanometer sizes for a wide range of technological purposes.  However, there is a fundamental limit to miniaturisation since small engines are {\it not} simply re-scaled versions of their larger counterparts.
 If the work performed during the duty cycle of any machine is comparable to thermal energy per degree of freedom, then one can expect that the machine will operate in ``reverse" over short time scales.  That is, heat energy from the surroundings will be converted into useful work allowing the engine to run backwards.  For larger engines, we would describe this as a violation of the Second Law of thermodynamics, as entropy is consumed rather than generated.  Until recently, this received little attention in the nanotechnology literature, as there was no theory capable of describing the probability of entropy consumption in such small engines. 

In the last fifteen years, several fluctuation theorems have been proposed that revolutionise our understanding and use of thermodynamics.  Firstly these new theorems lift the requirement of the thermodynamic limit.  This allows thermodynamic concepts to be applied to finite, even small systems.  
Secondly, these new theorems can be applied to systems that are arbitrarily far from equilibrium.  Thirdly for the first time, these theorems explain how macroscopic irreversibility appears naturally in systems that obey time reversible microscopic dynamics.  Resolution of the Loschmidt (Irreversibility) Paradox had defied our best efforts for more than 100 years. 

One of these fluctuation theorems, the Evans-Searles Fluctuation Theorem (Evans-Searles FT) \cite{Evans-PRE-1994, Evans-AP-2002, Searles-AJC-2004}, results in a generalisation of the Second Law of Thermodynamics so that it applies to small systems, including those that evolve far from equilibrium.  Another, the Crooks Fluctuation Theorem (Crooks FT) \cite{Crooks-JSP-1998, Crooks-PRE-1999}  provides a method of predicting equilibrium free energy difference from experimental information taken from nonequilibrium paths that connect two equilibrium states.  This FT can be used to derive the well known Jarzynski Equality \cite{Jarzynski-PRL-1997, Jarzynski-PRE-1997,Jarzynski-JSP-2000, Jarzynski-JSM-2004,Jarzynski-PTPS-2006, Crooks-PRE-2007}, which expresses the free energy difference between two equilibrium states in terms of an average over {\it irreversible} paths.  Both FTs are at odds with a traditional understanding of 19th century thermodynamics.  
Nevertheless, these theorems will be essential for the application of thermodynamic concepts to nanotechnology systems which are currently of such interest to biologists, physical scientists and engineers.

\subsection{The Evans-Searles Fluctuation Theorem}

In many areas of physical chemistry, researchers strive to understand new systems through deterministic equations of motion.  They seek to quantify microscopic forces and understand how a system responds to external perturbations, using techniques such as Molecular Dynamics simulation.  At the heart of this endeavour is the notion that if the equations of motion or trajectories of the system are known, then any question about that system may be answered.  However, such deterministic equations (such as Newton's equations) are time-reversible, so that for every trajectory there exists a conjugate, time-reversed trajectory or ``anti-trajectory"  which is also a solution to the equations.   The relative probabilities of observing bundles of conjugate trajectories can be used to quantify the ``macroscopic reversibility" of the system:  if the probability of observing all trajectories and their respective anti-trajectories are equal, the system is said to be reversible; on the other hand, if the probability of observing anti-trajectories is vanishingly small, we say that the system is irreversible.
The Second Law of Thermodynamics stipulates that a system evolves irreversibly in one ``time-forward" direction, {\it i.e.}, the probability of all anti-trajectories is zero.  However, the Second Law strictly applies to large systems or over long time scales and does not describe the reversibility of small systems that are of current scientific interest, such as protein motors and nano-machines.  This long-standing question of  how irreversible macroscopic equations, as summarised by the Second Law of Thermodynamics, can be derived from reversible microscopic equations of motion was first noted by Loschmidt\cite{Loschmidt-1876, Cercignani-1998} in 1876 and has been a paradox since then.  Boltzmann and his successors have simply side-stepped this issue with Boltzmann stating  ``as soon as one looks at bodies of such small dimension that they contain only very few molecules, the validity of this theorem [the Second Law of Thermodynamics] must cease" \cite{Broda-1983}

The Fluctuation Theorem (FT) of Evans \& Searles \cite{Evans-PRE-1994, Evans-AP-2002, Searles-AJC-2004} describes how a finite sized system's irreversibility develops in time from a completely time-reversible system at short observation times, to an irreversible one at long times.  It also shows how irreversibility emerges as the system size increases.  That is, it bridges the microscopic and macroscopic descriptions, relating a system's time-reversible equations of motion to the Second Law, and provides a quantitative resolution to the long-standing irreversibility paradox.   Specifically, the FT relates the relative probabilities, $p$, of observing trajectories of duration $t$  characterised by the dissipation function, $\Omega_t$, taking on arbitrary values ${\cal A}$ and ${-\cal A}$, respectively:
\begin{equation}
\frac{p(\Omega_t = {\cal A})}{p(\Omega_t=-{\cal A})} = \exp{({\cal A})}.
\label{eqn:FT}
\end{equation}
It is an expression that describes the asymmetry in the distribution of $\Omega_t$ over a particular ensemble of trajectories.  The dissipation function, $\Omega_t$,  is, in general, a dimensionless dissipated energy, accumulated along the system's trajectory; expressions for $\Omega_t$ differ from system to system.  However, any trajectory of the system that is characterised by a particular value $\Omega_t= {\cal A}$ has, under time-reversible mechanics, a conjugate or time-reversed anti-trajectory' with $\Omega_t = -{\cal A}$.  In this way, the LHS of the FT has also been interpreted as a ratio of the probabilities of observing trajectories to their respective anti-trajectories.  The dissipation function, $\Omega_t$, is an extensive property, {\it i.e.}, its magnitude scales with system size, and it also scales with the observation time, $t$.   Thus, eqn~\ref{eqn:FT} also shows that as the system size gets larger or the observation time gets longer, anti-trajectories become rare and it becomes overwhelmingly likely that the system appears time-irreversible, in accord with the Second Law.   That is, the evolution of a large macroscopic system proceeds preferentially in one direction.    In addition, eqn~\ref{eqn:FT} also shows that the ensemble average of the dissipation function is positive for all $t$ for all nonequilibrium systems and for any system size; {\it i.e.}, $\langle \Omega_t \rangle \ge 0$, \cite{Searles-AJC-2004}. We will refer to this as the Second Law Inequality.

\subsection{The Crooks Fluctuation Theorem and the Jarzynski Equality}

From classical thermodynamics, the work done by an external field to drive a system from one equilibrium state to another equilibrium state is equivalent to the change of free energy, $\Delta F$, between the states, only in the special case where the path is traversed quasi-statically.  
That is the path between the two states must be traversed so slowly that intermediate, as well as the initial and final states of the system, are all in thermodynamic equilibrium.  

Crooks' Fluctuation Theorem (Crooks FT)  \cite{Crooks-JSP-1998, Crooks-PRE-1999}  states something quite remarkable.  In the case of paths that are traversed at arbitrary rate, ranging from quasi-static to ``far-from-equilibrium", the distribution of trajectories, characterised by the work done by the external field over the duration of the trajectory, follows
\begin{equation}
\frac{p_f(W = {\cal A})}{p_r(W=-{\cal A})} = \exp{[\beta({\cal A}-\Delta F)]}.
\label{eqn:Crooks}
\end{equation}
where $\beta=1/(k_BT)$, $k_B$ is Boltzmann's constant and $T$ is the initial temperature of the system on which the external field does work, or equivalently the temperature of the surroundings with which the system is initially at equilibrium.  This expression is similar to Evans-Searles FT in that it relates distributions of trajectories, characterised by an energy, specifically the work, $W$.  While eqn~\ref{eqn:FT} describes the asymmetry in the distribution of trajectories starting from the same initial distribution, Crooks FT, eqn~\ref{eqn:Crooks}, relates trajectories initiated from two different equilibrium states, $A$ and $B$. That is, it considers (i) a distribution, $p_f$, of {\it forward} trajectories, $A \rightarrow B$, where the free energy change between equilibrium states A and B is $\Delta F = F_B-F_A$, and (ii) the distribution, $p_r,$ of {\it reverse} trajectories, $B\rightarrow A,$ where the respective equilibrium free energy change is $-\Delta F$.   Like the FT, Crooks FT also quantifies how irreversibility evolves out of reversible equations of motion.  A perfectly reversible (quasi-static) system is one where the work required to traverse $B \rightarrow A$ is equal but opposite in sign to the work required in the time-reversed trajectory, $A \rightarrow B$. Thus the RHS of eqn.~\ref{eqn:Crooks} is unity for these reversible paths and $W=\Delta F$, in agreement with classical thermodynamics.  

Taking the ensemble average of $-\beta W$ and using the Crooks FT gives, 
 \begin{equation}
\exp{(-\beta \Delta F)} = \langle \exp{(-\beta W)}\rangle_{f}.
\label{eqn:Jarzynksi}
\end{equation}
Here the notation $ \langle ...\rangle_{f}$ implies an ensemble average using the distribution function of state A, and the work is measure over forward trajectories $A \rightarrow B$.  This expression was first posed by Jarzynski in 1997 \cite{Jarzynski-PRL-1997, Jarzynski-PRE-1997} before eqn~\ref{eqn:Crooks} was discovered, and is known as the Jarzynski Equality.  It states that the free energy can be determined by measuring the work, $W$, done by an external field along dynamical paths that connect the two states. These forward paths may be traversed {\it at arbitrary rates}, so that the intervening states may not be in thermodynamic equilibrium.   This provides a completely new way of treating thermodynamics.  If instead of averaging the work, you average the exponential of the work, then you can calculate the equilibrium free energy difference from information obtained along {\it nonequilibrium} paths.  
On the practical side, Eqn~\ref{eqn:Jarzynksi}, suggests that measuring work on small microscopic processes could yield thermodynamic quantities $\Delta F$ that are traditionally inferred by calorimetric measurements.  The importance here is that in order to understand molecular-scale processes, it is necessary to probe them using molecular time/length scales.  

\subsection{Organisation of the Review}
In the following section we will introduce some background concepts required for understanding the derivation and consequences of the FTs.  We introduce general equations of motion for nonequilibrium systems and give expressions for the work and heat transferred from the systems. We discuss how thermostatting is achieved in simulations, and in particular discuss the thermostatting mechanisms that produce a canonical equilibrium distribution, since it is the canonical state that was originally treated in the Crooks FT.  In Section 3 we review the concepts of the original FT derivation made by Evans and Searles for systems governed by deterministic mechanics.  Today, many proofs exist for the FTs, extending to systems described by quantum, {\it e.g.} \cite{Jarzynski-PRL-2004, Esposito-PRE-2006} and stochastic dynamics {\it e.g.} \cite{Searles-PRE-1999, Lebowitz-JSP-1999}.  Here we limit our review to the original deterministic treatment and to concepts central to the understanding the FT in the context of modern statistical mechanics and its applications.  We extend this description to provide a derivation of the Crooks FT for deterministic systems so as to demonstrate the similar basis it has with the Evans-Searles FT, and we refer collectively to these as the FTs.  These derivations consider the response of equilibrium systems to external perturbations.  However, we also show how steady state fluctuation relations can be derived for systems that approach a unique steady state, and in those cases, how the FTs can be used to obtain Einstein and Green-Kubo relations.   In Section 4 we review applications of the FTs to experimental and model systems.

\section{\label{sec:basics} Background concepts}

The degrees of freedom for a system of particles can
be the represented by the vectors of time-dependent
generalized coordinates, ${\bf q}$, and momenta, ${\bf
p}$. In addition, a single point in the system's
phase-space is denoted as ${\bf \Gamma} \equiv ({\bf q}, {\bf
p})$. Consider a closed system that is in thermal
equilibrium with a reservoir. From equilibrium statistical mechanics,
we know that the equilibrium probability distribution
of the system is given by the canonical distribution
function,
\begin{equation}
f_{eq}({\bf\Gamma}) = \frac{e^{-\beta {\cal H}
({\bf\Gamma})}}{Z} \,, \label{eqn:canonical}
\end{equation}
where $Z = \int e^{-\beta {\cal H}} \, d\bf{\Gamma} $
is the equilibrium partition function, 
and $\cal H$ is the internal 
energy\footnote{More exactly, $\cal H$ is the phase-variable
corresponding to the internal energy. For simplicity,
we will use the term ``internal energy" to refer to
$\cal H$, and will refer to the thermodynamic internal
energy $U = \langle {\cal H} \rangle$ as the ``mean
internal energy".} which is the sum of  the kinetic and potential energies, $K({\bf p})$
and $\phi ({\bf q}),$ of the system.

A closed \emph{adiabatic} system can exchange energy
with its environment in the form of work. We can think
of work as the form of energy exchange that is
directly controllable by the environment. For example,
it might be desirable to change the mean internal
energy, $U = \langle {\cal H} \rangle,$ of the system,
and this can be achieved by externally controlling
some parameter $\lambda$ in the potential energy
function of the system, $\phi$. Examples of such
$\lambda$-parameters include the switch on an externally applied
electric field in a crystalline salt, the
trapping constant in an optical trap holding a
colloidal particle, or a mathematical agent that
changes the size of Lennard-Jones spheres in a
computer simulation. We emphasise this mode of
external control by formally making $\lambda$
time-dependent and by writing the internal energy as
\begin{equation}
{\cal H}({\bf \Gamma}, s)= K({\bf p}) + \phi({\bf q},
\lambda(s))\,. \label{eqn:fullHam}
\end{equation}
When an external agent does work on a system without changing its underlying equilibrium state, which has mean internal energy $U$, we refer to that field as a purely dissipative field, denoting it generally by ${\bf F}_e$. This dissipative field does not figure in the underlying equilibrium distribution
or partition function, but drives the system away from
equilibrium.  While it may be possible to represent the external agent using either ${\bf F}_e$ or $\lambda$, we choose the convention that  if $\dot{\lambda}=0$ and ${\bf F}_e=0$, the system will always relax to an (nondissipative) equilibrium state, and that if ${\bf F}_e \ne 0$, the system will never relax to a nondissipative state. This distinction may depend on the state of system ({\it e.g.,} fluid or solid). Examples of such dissipative fields include a fluid under a shear flow, dragging a colloidal particle in a fluid,
and an electric field acting on a molten salt. Under
adiabatic conditions ({\it i.e.} the rate of heat exchange with the reservoir is $\dot{Q}=0$), the combined action of both
kinds of external agents, {\it i.e.,} a time dependent potential represented by a $\lambda$-parameter and a dissipative field ${\bf F}_e$, results in the equations of motion,  
\begin{eqnarray}
\dot{\bf q} &= & \frac{\partial{\cal H}({\bf \Gamma},s)}{\partial {\bf p}} + {\bf C}({\bf \Gamma}) \cdot {\bf F}_e(s) \nonumber \\
\dot{\bf p} &= & -\frac{\partial{\cal H}({\bf
\Gamma},s)}{\partial {\bf q}} + {\bf D}({\bf \Gamma})
\cdot {\bf F}_e(s) \,,\label{eqn:adiabaticeom}
\end{eqnarray}
with ${\cal H}({\bf \Gamma},s)$ given by
eqn.~\ref{eqn:fullHam}.\footnote{We must point out
here that the notation does not imply that ${\bf F}_e$
is a ``force", and neither does it have to always be a
vector. For example, it could be a second order
tensor, such as the velocity gradient tensor in a
fluid. The coupling tensors ${\bf C}$ and ${\bf D}$
are functions of $\bf \Gamma$ and have no explicit
time-dependence, and are formally one tensorial order
higher than ${\bf F}_e$.}

For an externally driven adiabatic system, the rate of increase of
$\cal H$ must be identically equal to the rate of work
$\dot{W}$ done on the system by the environment. Thus,
\begin{equation}
\dot{W}({\bf \Gamma}, s) = \dot{\cal H}^{ad}({\bf \Gamma}, s)  = \dot{\lambda}\frac{\partial {\cal H}({\bf \Gamma},s)}{\partial \lambda} + \dot{\bf q}\cdot \frac{\partial {\cal H}({\bf \Gamma},s)}{\partial {\bf q}} + \dot{\bf p}\cdot \frac{\partial {\cal H}({\bf \Gamma},s)}{\partial {\bf p}} \,,\\
\end{equation}
where the superscript {\it ad} emphasises adiabatic conditions.
Using eqn.~\ref{eqn:adiabaticeom}, we obtain,
\begin{equation}
\dot{W}({\bf \Gamma}, s) =  \dot{\lambda}
\frac{\partial \phi({\bf q},\lambda(s))}{\partial
\lambda} - V {\bf J}({\bf \Gamma}) \cdot {\bf
F}_e(s) \,, \label{eqn:working}
\end{equation}
where $V$ is the volume of the system, and ${\bf J({\bf \Gamma})}$, the dissipative flux due to
the field ${\bf F}_e(s)$  is formally defined through
the equation:
\begin{equation}
 V {\bf J} \cdot {\bf F}_e \equiv  - \left(\frac{\partial {\cal H}}{\partial {\bf q}} \cdot {\bf C} \cdot  {\bf F}_e   +  \frac{\partial {\cal H}}{\partial {\bf p}}\cdot {\bf D} \cdot {\bf F}_e \right)\,.\\
\end{equation}

In the case of small systems such as protein motors or
artificial nanomachines, it is quite difficult to thermally isolate the system to
achieve perfectly adiabatic conditions. Moreover, in
most applications of interest, such systems typically
function in an environment of constant temperature.
Molecular dynamics simulations of small systems have
employed ``thermostats" which involve appending
eqn~\ref{eqn:adiabaticeom} with a mathematical
constraint to fix the temperature $T$. For example,
with a Gaussian isokinetic thermostat,
eqns.~\ref{eqn:adiabaticeom} take the form
\cite{EvansMorriss}:
\begin{eqnarray}
\dot{\bf q} &= & \frac{\partial{\cal H}({\bf \Gamma},s)}{\partial {\bf p}} + {\bf C}({\bf \Gamma}) \cdot {\bf F}_e(s) \nonumber \\
\dot{\bf p} &= & -\frac{\partial{\cal H}({\bf
\Gamma},s)}{\partial {\bf q}} + {\bf D}({\bf \Gamma})
\cdot {\bf F}_e(s) - \alpha({\bf \Gamma}) {\bf S}
\cdot {\bf p}. \label{eqn:fulleom}
\end{eqnarray}
Here, $\alpha$ is a thermostat multiplier\footnote{In the case of a Nos\'e Hoover thermostat\cite{Hoover-PRA-1985}, $\alpha$ becomes an additional independent variable and is a function of time governed by an additional equation rather than a direct function of $\bf \Gamma$ \cite{EvansMorriss}.}, and ${\bf
S}$ is a diagonal matrix (with 1's and 0's on the
diagonal) that describes which components of the
system are thermostatted\footnote{When the external field is removed, ${\bf F_e}=0$, and the potential's parameter is held fixed, $\dot{\lambda}=0$, the equilibrium these equations, eqn \ref{eqn:fulleom}, will relax to is given by $f_{eq}({\bf \Gamma})=\frac{\exp(-\beta \mathcal{H}({\bf \Gamma}, \lambda))}{Z}\delta ({\bf p\cdot S\cdot p}-2mK)$, where the Dirac delta function accounts for the kinetic energy of the thermostatted particles being held fixed to the value $K$ and $m$ is the particle mass. The partition function is given by $Z=\int d{\bf \Gamma}\; \exp(-\beta \mathcal{H}({\bf \Gamma}, \lambda))\delta ({\bf p \cdot S \cdot p}-2mK)$.}. Several other mathematical
constraints can be constructed, all of which may be
argued to be ``artificial". We will discuss the
implications of such artificial thermostats, and the
constraints they must satisfy shortly.

As the system is closed, an increase in the
internal energy must equal the sum of work done on the
system by the environment and the heat added to the
system by the thermostat. However, it is clear that the functional form
of the expression for the rate of work must still be
given by the relation in eqn.~\ref{eqn:working} regardless of whether the system is thermostatted or not. 
From the First Law of Thermodynamics, the expression for
the rate of heat exchange is $\dot{Q} = \dot{\cal H} - \dot{W}$,
with $\dot{W}$ given by eqn.~\ref{eqn:working}. The
actual expression for the rate of change of $Q$ depends on the
mathematical form of the thermostat. For the
thermostat represented in eqn.~\ref{eqn:fulleom},
\begin{equation}
\dot{Q}({\bf \Gamma}, s) = - \alpha({\bf \Gamma})
\frac{\partial {\cal H}}{\partial {\bf p}}\cdot {\bf
S} \cdot {\bf p} \,. \label{eqn:heating}
\end{equation}
The total work done on a closed system is hence,
\begin{equation}
W = \int_0^t  \left(\dot{\lambda} \frac{\partial
\phi({\bf q},\lambda(s))}{\partial \lambda} - V {\bf
J}({\bf \Gamma},s) \cdot {\bf F}_e(s) \right)\, d s
\,, \label{eqn:work}
\end{equation}
and the heat added to the system is $Q = - \int_0^
t \dot{Q} \, d s$.  We note here that the integrals
above are path integrals. However, since the dynamics
described by eqn.~\ref{eqn:fulleom} are completely
deterministic, $W$ and $Q$ are functions solely of the
initial point in phase-space at $s = 0$, and the
duration $t$. That is, $W = W({\bf \Gamma}_0, t)$ and
$Q = Q({\bf \Gamma}_0, t)$.

Consider a particular trajectory initiated at time
$s=0$ at ${\bf \Gamma}_0 \equiv ({\bf q}_0,{\bf
p}_0)$,  that terminates after time $t$ at ${\bf
\Gamma}_t \equiv ( {\bf q}_t, {\bf p}_t)$. Let
$d\,{\bf \Gamma}_s$ represent an infinitesimal volume
of phase space at time $s$ about the point ${\bf
\Gamma}_s$. As the dynamics is deterministic, the
trajectory is completely determined by the phase space
coordinates at any time $s$ along the trajectory, and
the duration or observation time, $t$, of the
trajectory. Consequently, for every initial state
within a volume element $d{\bf \Gamma}_0$ there exists
a unique destination point within volume element
$d{\bf \Gamma}_t$. As the trajectories in an infinitesimal bundle around the initial state, $d{\bf \Gamma}_0$, form the later bundle $d{\bf \Gamma}_t$, the ratio of the volumes of
the infinitesimal volume elements vary as
\begin{equation}
\left| \frac{d\,{\bf \Gamma}_t}{d\,{\bf \Gamma}_0} \right|
= \frac{\delta V({\bf \Gamma}_t)}{\delta V({\bf \Gamma}_0)} 
=\exp{\left(\int_0^t \, \Lambda({\bf \Gamma}_s)\,
ds\right)}. \,, \label{eqn:phasevolcomp}
\end{equation}
Here $d\,{\bf \Gamma}_t / d\,{\bf \Gamma}_0$ is the
Jacobian of the transformation of the initial
 ${\bf \Gamma}_0$ to the final ${\bf \Gamma}_t$, and
\begin{equation}
\Lambda({\bf \Gamma}_s,s) \equiv
\frac{\partial}{\partial {\bf \Gamma}_s} \cdot
\dot{\bf \Gamma}_s = \left(\frac{
\partial }{\partial {\bf q}} \cdot \dot{\bf q} +
\frac{
\partial }{\partial {\bf p}} \cdot \dot{\bf
p}\right)_s
\,, \label{eqn:lambdadef}
\end{equation}
is the \emph{phase-space compression factor}. It is
noted that the integral on the right-hand side of
eqn.~\ref{eqn:phasevolcomp} is also a path-integral.

Equations~\ref{eqn:phasevolcomp} and
\ref{eqn:lambdadef} show how the volume of a small
region of phase-space changes as it evolves in time.
For adiabatic systems, there is no change of
phase-space volume along a trajectory, and we require
that $\Lambda^{ad} = 0$. From the equations for an
adiabatic system (eqn.~\ref{eqn:adiabaticeom}), and
the definition of $\Lambda$ above, we see that the
field ${\bf F}_e$ and the coupling tensors $\bf C$ and
$\bf D$ must be such that
\begin{equation}
\frac{\partial }{\partial {\bf q}} \cdot {\bf C} \cdot
{\bf F}_e = \frac{
\partial }{\partial {\bf p}} \cdot {\bf D}
\cdot {\bf F}_e = 0 \,, \label{eqn:AIG}
\end{equation}
irrespective of whether the system is thermostatted or
not. This condition is known as the \emph{Adiabatic
Incompressibility of Phase-Space} (AI$\bf \Gamma$). 
However, for thermostatted systems in a driven steady state, 
a contraction of phase space occurs continually, as the
initial phase volume shrinks to a fractal attractor of
lower dimension than the ostensible phase-space. 
For appropriately selected thermostats\footnote{
These include the Gaussian isokinetic
thermostat and the Nos\'e-Hoover thermostat. However, we
note that eqn.~\ref{eqn:lambdadef} has to be extended
in the case of the Nos\'e-Hoover thermostat as detailed
in Ref. \cite{EvansMorriss}.}, the 
phase-space contraction factor is directly proportional to
the rate of heat exchange with the thermostat \cite{Williams-PRE-2004, Bright-JCP-2005},
\begin{equation}
\dot{Q}({\bf \Gamma}, s) = k_B T \,\Lambda({\bf
\Gamma}, s) \,. \label{eqn:thermostatcondn}
\end{equation}

Since the same exclusive set of trajectories passes
through both the phase volumes $d{\bf \Gamma}_0$ and
$d{\bf \Gamma}_t$, the differential probability
\emph{measures} of the two infinitesimal volumes must
be identical:
\begin{equation}
dP \,(d{\bf \Gamma}_0, 0)= dP\,(d{\bf \Gamma}_t, t)
\,. \label{eqn:dPequality}
\end{equation}
We can express the probability measure $P$ in terms of
the probability density $f$ as
\begin{equation}
dP\,(d{\bf \Gamma}_s, s) = f({\bf \Gamma}_s, s)\,\delta V({\bf
\Gamma}_s)\,, \label{eqn:fdef}
\end{equation}
where $f({\bf \Gamma}_s, s)$ is the time-dependent
phase space probability density. The observation that
the probability measure is conserved in phase-space
also leads to the Liouville equation for the
probability density:
\begin{equation}
\frac{ \partial f({\bf \Gamma}, s)}{\partial s} +
\dot{\bf q} \frac{ \partial f({\bf \Gamma},
s)}{\partial {\bf q}} + \dot{\bf p} \frac{ \partial
f({\bf \Gamma}, s)}{\partial {\bf p}} = -f({\bf
\Gamma}, s) \Lambda({\bf \Gamma},s) \,.
\label{eqn:liouville}
\end{equation}
We can recast eqn.~\ref{eqn:liouville} into the
following Lagrangian form:
\begin{equation}
\frac{ d \ln f({\bf \Gamma}, s)}{d s} = - \Lambda({\bf
\Gamma},s) \,, \label{eqn:liouville2}
\end{equation}
from which it can be shown that
\begin{equation}
f({\bf \Gamma}_t, t) = f({\bf \Gamma}_0,0) \,
\exp{\left(-\int_0^t \Lambda({\bf \Gamma}_s,s)\,ds
\right)}.
\end{equation}
This equation is also obtained directly from
eqns.~\ref{eqn:phasevolcomp}, \ref{eqn:dPequality},
and \ref{eqn:fdef}.

One may enquire about the effect of the introduction of fictitious thermostats 
in eqn \ref{eqn:fulleom} and the possible introduction of artifacts \cite{EvansMorriss, Klages-2007}. As mentioned above, eqn \ref{eqn:thermostatcondn} ensures that the
equations of motion correctly sample the appropriate
equilibrium distribution function in the absence of
${\bf F}_e$ and when $\dot{\lambda} = 0$. We also know that these equations of
motion do not introduce artifacts when used to
determine the linear response of a system to a small
external field ${\bf F}_e$. Further, the equilibrium
correlation functions used in the Green Kubo integrals
are affected by the thermostat at most as
$\mathcal{O}(1/N)$ where $N$ is the number of
particles. A general description of a system that is
driven past a linear response is difficult, and in the
nonlinear regime, synthetic thermostatted dynamics as
in eqn.~\ref{eqn:fulleom} may produce artifacts.
Moreover, Mori-Zwanzig theory
\cite{Zwanzig-2001,EvansMorriss} can no longer be
combined with the Onsager regression hypothesis
\cite{EvansMorriss} to rigorously derive stochastic
equations. To address this, we can arrange things such
that the thermostat only acts on particles that are in
a region, which is spatially far enough removed from
the nonequilibrium process such that it remains in
local equilibrium
\cite{Ayton-JCP-2001,Evans-AP-2002,Williams-PRE-2004,Williams-MP-2007}.
A detailed theoretical and simulation study has shown
how this approach can, for an infinite family of
thermostats, result in the same behavior for the
system of interest \cite{Williams-PRE-2004}. Equations
of motion for isothermal-isobaric systems with a
thermostat and barostat which are external to the
system of interest, have also been developed
\cite{Williams-MP-2007}. These developments are
theoretically important because they allow the
derivation of important theorems which require the
condition of ergodic consistency, which will be
discussed shortly. For driven systems a satisfying treatment requires
a mechanism for heat exchange. The development of synthetic
thermostats, which only act on regions removed from the
system of interest, allows the ergodic consistency
condition to be satisfied without introducing
artifacts when far from equilibrium. This outcome is
not easy to arrive at by other means.

\section{Fluctuation Theorems from Deterministic
Dynamics}
\subsection{Evans-Searles Fluctuation Theorem}
As mentioned in the Introduction, the Evans-Searles FT
shows how irreversibility emerges naturally in systems
whose equations of motion are time-reversible. In
order to fully appreciate the substance of this
FT, we need to first define two fundamental concepts:
microscopic time-reversibility, and macroscopic
irreversibility.

\subsubsection{Microscopic Time-Reversibility}

The equations of motion in section \ref{sec:basics} describe the time
evolution of a point, $\bf \Gamma$, and may depend explicitly on the
time\footnote{That is the equations of motion may be nonautonomous.} due to
the possible time-dependencies of $\lambda$  and ${\bf F}_e$. If the
equations of motion are reversible, then there exists a time reversal
mapping that transforms the point ${\bf \Gamma} \equiv ({\bf q},{\bf p})$ to $
{\bf \Gamma}^\ast$ such that if we generate a trajectory starting at $
{\bf \Gamma}_0$ and terminating at ${\bf \Gamma}_t$, then under the same
dynamics, we start at ${\bf \Gamma}_t^\ast \equiv ({\bf \Gamma}_t)^\ast$ and
arrive back at ${\bf \Gamma}_0^\ast\equiv ({\bf \Gamma}_0)^\ast$ after time
t. We refer to a trajectory and its anti-trajectory as a conjugate pair of trajectories. 
The time-average of properties that are even under
the mapping will have equal values for the trajectory and its conjugate,
whereas the time-average of properties that are odd under the mapping will
have values with equal magnitude, but opposite signs for the trajectory and
its conjugate.  For many dynamics, ({\it e.g.} Newtonian dynamics), the
appropriate mapping gives ${\bf \Gamma}^*=({\bf q},-{\bf p})$.

For the equations of motion\footnote{In the case of Nos\'e Hoover equations
of motion we have an extra degree of freedom due to the thermostat
multiplier, $\alpha(t)$. This phase space variable must be reversed along
with the momentum upon applying the time reversal mapping ${\bf
\Gamma}^\ast$.}, eqn \ref{eqn:adiabaticeom} or \ref{eqn:fulleom}, to
satisfy this condition we must have\footnote{Often there is an external
symmetry, e.g. for a fluid we may be able to drive a process in the
opposite direction and consider it equivalent to the original direction,
such that we may consider the conditions ${\bf C} ( {\bf \Gamma} ) \cdot
{\bf F}_e(s) = {\bf C} ( {\bf \Gamma}^\ast) \cdot {\bf F}_e(t-s)$ and ${\bf
D} ( {\bf \Gamma} ) \cdot {\bf F}_e(s) = -{\bf D} ( {\bf \Gamma}^\ast)
\cdot {\bf F}_e(t-s)$ to in effect provide time reversal symmetry. This is
why the Evans-Searles FT can often allow protocols which have an odd time parity.}
\begin{eqnarray}
\phi({\bf q}, \lambda(s)) &=& \phi({\bf q},
\lambda(t - s)) \,,\nonumber\\
{\bf C} ( {\bf \Gamma} ) \cdot {\bf F}_e(s) &=& -{\bf C} ( {\bf
\Gamma}^\ast) \cdot {\bf F}_e(t-s) \,,\nonumber\\
{\bf D} ( {\bf \Gamma} ) \cdot {\bf F}_e(s) &=& {\bf D} ( {\bf
\Gamma}^\ast) \cdot {\bf F}_e(t-s) \,.
\label{eqn:evenparity}
\end{eqnarray}

\begin{figure}
\includegraphics[scale=0.8]{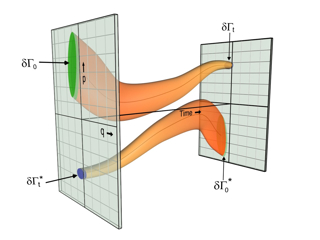}
\caption{An illustration of a set of neighbouring deterministic  trajectories of duration time, $t$, (top tube) and their corresponding set of time-reverse or ``anti-trajectories" (lower tube) in coordinate, momentum ${\bf \Gamma}\equiv\{{\bf q},{\bf p}\}$ and time, $s$, space where an external agent does work on the system.   This external agent, either $\lambda$-parameter in the potential energy, or a purely dissipative field, ${\bf F}_e$  incorporated in the equations of motion, must have even time-parity for the dynamics to remain time-reversible.  For every trajectory that starts at ${\bf \Gamma}_0=\{ {\bf q}_0,{\bf p}_0 \}$ in the volume element $\delta{\bf \Gamma}_0$ and ends at ${\bf \Gamma}_t=\{ {\bf q}_t,{\bf p}_t \}$ in volume element $\delta{\bf \Gamma}_t$ at some time $t$ later, there exists the anti-trajectory, whose coordinates, at any time $s$ along the trajectory starting at $s=0,$ are given by ${\bf \Gamma}^\ast_{t-s} = \{ {\bf q}_{t-s}, -{\bf p}_{t-s} \}$.
Thus, the anti-trajectory starts at ${\bf \Gamma}_t^\ast \equiv \{{\bf q}_t,-{\bf p}_t \}$ in volume element $\delta{\bf \Gamma}^\ast_t$ and terminates after a time $s= t$ at ${\bf \Gamma}_0^\ast \equiv \{ {\bf q}_0, -{\bf p}_0 \}$  in $\delta{\bf \Gamma}_0^\ast.$  For thermostatted systems, if the action of the external agent does work on the system,  there is a contraction of phase-space volume in time, {\it i.e.,} $\delta{\bf \Gamma}_t < \delta{\bf \Gamma}_0$ as represented in the figure by the shrinking of the tube's cross sectional area in time.  As the equations of motion are time-reversible, the phase-space volume increases from $s=0$ to $s=t$ and the size of the volume elements $\delta{\bf \Gamma}_t$ and $\delta{\bf \Gamma}_t^\ast$ are equal.}
\label{fig:tubes}
\end{figure}

Let us now consider a system of particles whose
overall equations of motion are time-reversible. As
discussed above, for every trajectory that is
initiated at ${\bf \Gamma}_0$ and terminates at ${\bf
\Gamma}_t $ in a system with microscopically
time-reversible dynamics, there exists a unique
anti-trajectory that starts at the phase-space point
${\bf \Gamma}^\ast_t$ at $s = 0$ and ends at ${\bf
\Gamma}^\ast_0$ at $s = t$. The bundle of
anti-trajectories  at time $t$ passes through the
volume element $d{\bf \Gamma}^\ast_0$ centered about
the point ${\bf \Gamma}^\ast_0$.  However, the size of
the volume element $d {\bf \Gamma}^\ast_t$ is equal to
that of $d{\bf \Gamma}_t$. Moreover, if there is a
volume contraction from $d{\bf \Gamma}_0$ to $d{\bf
\Gamma}_t$ as shown in Figure~\ref{fig:tubes}, then
there is an equivalent volume expansion associated
with the bundle of anti-trajectories.

Given a system whose equations of motion are
microscopically time-reversible, is the
macroscopically observed behaviour reversible as well?
As Kelvin and Loschmidt pointed out in the 1870's, because Newtonian
equations of motion are microscopically
time-reversible, for every trajectory there is a
anti-trajectory which is also a solution to the
equations. One might then conclude that microscopically time-reversible
systems must also be macroscopically reversible.
However the Second Law of Thermodynamics stipulates that a
macroscopic system evolves overwhelmingly in one, time-forward
direction and is ``irreversible". The question of how
microscopically time-reversible dynamics gives rise to
observable macroscopic irreversibility, is indeed
``Loschmidt's Paradox''. To resolve this paradox, we first
need an unambiguous measure of ``macroscopic
irreversibility'', that is consistent with
classical thermodynamics in the thermodynamic limit,
and applies to microscopic time-reversible equations of
motion.

\subsubsection{Macroscopic irreversibility: the
Dissipation Function}%
A system is said to undergo a \emph{macroscopically
reversible} process in the time interval  $0 \leq  s
\leq t$, if
\begin{enumerate}
\item the system is \emph{ergodically consistent}.
That is for every trajectory that initiates at ${\bf
\Gamma}_0$, the starting coordinates of its respective
anti-trajectory, ${\bf \Gamma}^\ast_t$ is represented
in the phase space of the system at $s=0$, or
equivalently, the probability density of the initial
coordinates of anti-trajectories at time $s=0$ is
non-zero:   $f({\bf \Gamma}^\ast_t,0) \ne 0$, for all
${\bf \Gamma}_0$.%
\item the probability of observing any bundle of
trajectories, occupying an infinitesimal volume, is equal 
to the probability of observing the conjugate bundle 
of anti-trajectories, or
\begin{equation}
dP(d{\bf \Gamma}_0, 0) = dP(d{\bf \Gamma}^\ast_t,
0)\,. \label{eqn:macrocondition}
\end{equation}
\end{enumerate}

The latter condition for macroscopic reversibility can
be written more conveniently in terms of the
distribution function of the phase space:
\begin{eqnarray}
f({\bf \Gamma}_0, 0)d V({\bf \Gamma}_0) = f({\bf
\Gamma}^\ast_t, 0) d V({\bf \Gamma}^\ast_t)\,, \nonumber
\end{eqnarray}
or,
\begin{eqnarray}
 \frac{ f({\bf
\Gamma}_0, 0)}{f({\bf \Gamma}^\ast_t, 0)} 
\left| \frac{d{\bf \Gamma}_0}{d{\bf \Gamma}^\ast_t} \right| = 1\,.
\nonumber
\end{eqnarray}
But, as mentioned earlier, the volume of $d{\bf
\Gamma}^\ast_t$ is the same as $d{\bf \Gamma}_t$.
Hence, from eqn.~\ref{eqn:phasevolcomp}, we see that
eqn~\ref{eqn:macrocondition}, a condition for macroscopic reversibility,  becomes
\begin{equation}
\ln\left[ \frac{ f({\bf \Gamma}_0, 0)}{f({\bf
\Gamma}^\ast_t, 0)} \right] - \int_0^t \Lambda({\bf
\Gamma}_s, s) \,ds \,=\, 0\,,
\label{eqn:condition}
\end{equation}
for any initial coordinate ${\bf \Gamma}_0$. Indeed, a quantitative measure of
\emph{irreversibility} associated with a system with
microscopically time-reversible dynamics over the
interval $0 \le s \le t$, $\Omega_t$  may be defined as the inequivalence of eqn~\ref{eqn:condition}
\begin{eqnarray}
\Omega_t({\bf \Gamma}_0) & = &  \ln{\left[    \frac{ d P(d{\bf \Gamma}_0, 0)}{d P(d{\bf \Gamma}^\ast_t, 0)}  \right] }\,\nonumber\\
& = &  \ln{ \left[  \frac{f({\bf \Gamma}_0,0)}{f({\bf
\Gamma}^\ast_t,0)}  \right]  } - \int_0^t \Lambda({\bf
\Gamma}_s, s)\, ds\,.\label{eqn:Omega1}
\end{eqnarray}
The dissipation function $\Omega_t$ is completely
determined for a deterministic trajectory by the initial coordinate, ${\bf \Gamma}_0$, and the duration of the trajectory, $t$. We note here that the time-reversibility of the
dynamics dictates  that conjugate pairs of trajectories are characterised by the same magnitude of $\Omega_t$, but of opposite sign:
\begin{eqnarray}
\Omega_t({\bf \Gamma}^\ast_t) = -\Omega_t({\bf
\Gamma}_0) \,. \label{eqn:posneg}
\end{eqnarray}
Furthermore, if $\Omega_t=0$ for all trajectories
initiated anywhere in phase-space, then the system is
in equilibrium and the probabilities of observing
any trajectory and its corresponding anti-trajectory
are equal. If $\Omega_t > 0$ for a trajectory, then
the corresponding anti-trajectory is less likely to be
seen, and if the ensemble average is greater than zero, $\left< \Omega_t \right> > 0$, we have macroscopic
dynamics moving in the ``forward direction". 
If $\left< \Omega_t \right> < 0$, then we would have
macroscopic dynamics in the reverse direction. Thus,
$\langle \Omega_t \rangle \ne 0$, 
is the condition for macroscopic irreversibility. Our
knowledge of the Second Law however seems to suggest
that the arrow of time points unambiguously in one
firm direction, accordingly
\begin{equation}
\langle \Omega_t \rangle \geq 0 \,.
\end{equation}
We explain how this comes about next.

\subsubsection{The Evans-Searles FT}

We consider trajectories of duration $t$  in phase-space by
selecting all those initial coordinates ${\bf \Gamma}_0$ for which
$\Omega_t$ takes on some value ${\cal A}$ between ${\cal A} \pm d{\cal A}$
 and thus obtain the probability density 
\begin{equation}
p(\Omega_t = {\cal A}) = \int d{\bf\Gamma}_0 \, \delta[\Omega_t({\bf
\Gamma}_0)-{\cal A}]f({\bf \Gamma}_0,0).
\label{eqn:forwardp}
\end{equation}
Upon recognising that ${\bf \Gamma}_0$ is merely a dummy variable of integration, we may write down the conjugate probability as,
\begin{equation}
p(\Omega_t = -{\cal A}) = \int d{\bf\Gamma}_t^\ast \, \delta[\Omega_t({\bf
\Gamma}_t^\ast)+{\cal A}]f({\bf \Gamma}_t^\ast,0).
\label{eqn:reversep}
\end{equation}
Note that eqn \ref{eqn:reversep} selects those trajectories which are conjugate to those which are selected by eqn \ref{eqn:forwardp}.
Using the definition of $\Omega_t$ in eqn \ref{eqn:Omega1} along with eqn \ref{eqn:posneg} and  eqn \ref{eqn:phasevolcomp}, we have
\begin{eqnarray}
p(\Omega_t = -{\cal A})&=&  \exp{[-{\cal A}]}\int
d{\bf \Gamma}_0 \, \delta[\Omega_t({\bf \Gamma}_0)-{\cal A})f({\bf
\Gamma}_0, 0),
\end{eqnarray}
which leads to the Evans-Searles Fluctuation Theorem
(Evans-Searles FT):
\begin{equation}
\frac{p({\Omega_t}={\cal A})} {p({\Omega_t}=-{\cal A})
}=\exp{[{\cal A}]}\,,
\end{equation}
and using this we can average over all values of ${\cal A}$ to give the Second Law Inequality, $\langle \Omega_t \rangle \geq 0$, 
\cite{Searles-AJC-2004}.  In the above derivation of the Evans-Searles FT it was assumed that :
\begin{enumerate}
\item{the dynamics is ergodically
consistent with the initial distribution function} 
\item{$f({\bf \Gamma},0) = f({\bf \Gamma}^\ast,0)$, and}
\item{the dynamics are deterministic
and microscopically reversible.}
\end{enumerate}
For systems of particles, the third condition implies
that the time-dependent $\lambda$ and ${\bf F}_e$ must
have an even time-parity $0 \le s \le t$. \footnote{If symmetry permits, an odd time parity is also acceptable.} These are sufficient conditions for the Evans-Searles FT to be valid, but the condition of microscopic reversibility can be relaxed to some degree, and stochastic versions of the Evans-Searles FT \cite{Searles-PRE-1999, Crooks-PRE-1999} exist.

What is the significance of the Evans-Searles FT?  One of the most important consequences of the Evans-Searles FT is that it shows how macroscopic irreversibility can eventuate from microscopically reversible equations of motion.  As described above, the systems we are considering are microscopically time-reversible.  The Evans-Searles FT defines the variable $\Omega_t$ which is a time-averaged phase variable and is zero for all initial phases if the system is macroscopically reversible, but will be non-zero for some initial conditions if it is macroscopically irreversible.  Furthermore, the Evans-Searles FT shows that $\left\langle \Omega_t \right\rangle \geq 0$ for any ergodically consistent, microscopically time-reversible system, and only zero if the system is at equilibrium. The significance of the Evans-Searles FT has been discussed in \cite{Evans-AP-2002, Searles-AJC-2004}, and other features are discussed in Sections 3.4 below

However, where does the irreversibility come from?  In obtaining the Evans-Searles FT, it is assumed that the initial distribution function is known and then, typically, the response of this system to a field, ${\bf F}_e$, or variation of $\lambda$ is considered.  Thus, we make the assumption of causality. If, instead, we had assumed that the system ends in a known state, we would have obtained the result $\left\langle \Omega_t \right\rangle \leq 0$.  Therefore the assumption of causality underlies the final result \cite{Evans-PRE-1996}.

\subsubsection{Dissipation Function for systems initially in a Canonical Ensemble}
Similar to the work, $W({\bf \Gamma}_0,t)$, the dissipation function can be expressed in terms of a trajectory of duration $t$ with initial coordinate ${\bf \Gamma}_0$, {\it i.e.} $\Omega_t({\bf \Gamma}_0)$.  Analogous to $W$, we restrict ourselves to trajectories initiated at equilibrium in the canonical ensemble and consider the action of both, a time-dependent $\lambda$-controlled potential where $\lambda(s=0) = A$ initially and $\lambda(s=t)=B$ finally, as well as a time-dependent dissipative field, ${\bf F}_e$.  For the definition of the dissipation function, eqn \ref{eqn:Omega1}, becomes

\begin{eqnarray}
\Omega_t({\bf \Gamma}_0) & = & \ln \frac{f_{eq} \,({\bf \Gamma}_0, \lambda =
A)}{f_{eq} \,({\bf \Gamma}_t^\ast, \lambda = A)}  -\int_0^t ds\, \Lambda({\bf \Gamma}_s,s) \\
&=& \beta \left[{\cal H} ({\bf \Gamma}_t^\ast, \lambda =
A)-{\cal H} ({\bf \Gamma}_0, \lambda = A) \right]  -\beta \int_0^t ds\, \dot{Q}({\bf \Gamma}_s,s).
\end{eqnarray}
Noting that the coordinates of the trajectory and anti-trajectory, ${\bf \Gamma}_s$ and ${\bf \Gamma}_s^\ast$, differ only in the direction of momenta, ${\bf p}$, and that ${\cal H}$ is even in momenta, the LHS can be cast as a time integral over the trajectory,
$$\Omega_t({\bf \Gamma}_0) = \int_0^t ds\, \bigg[ \dot{\cal H}({\bf \Gamma}_s, \lambda(s))-\dot{Q}({\bf \Gamma}_s, s) - \dot \phi({\bf q}, \lambda(s)) + \dot \phi({\bf q},\lambda=A)\bigg],$$
so that
\begin{eqnarray}
\Omega_t({\bf \Gamma}_0) & = &\int_0^tds \bigg[ \dot{\cal H}^{ad}({\bf \Gamma}_s, \lambda(s)) - \dot \phi({\bf q}, \lambda_s) + \dot \phi({\bf q},\lambda=A)\bigg] \\
& = & W({\bf \Gamma}_0, t) - \int_0^tds\bigg[\dot \phi({\bf q}, \lambda_s) + \dot \phi({\bf q},\lambda=A)\bigg] ,
\end{eqnarray}
or explicitly in terms of the potential and the external field as,
\begin{eqnarray}
\Omega_t({\bf \Gamma}_0) = \beta\int_0^t ds\bigg[-{\bf J}\cdot {\bf F}_e V + \dot{\bf q}\bigg(\frac{\partial \phi({\bf q},\lambda(s))}{\partial {\bf q}}- \frac{\partial \phi({\bf q},\lambda=A)}{\partial {\bf q}}\bigg)\bigg].
\end{eqnarray}

\subsection{Crooks FT for Deterministic Dynamics}
We can generate probability distribution functions for $W$, which is the work done on the system, be it in terms of a parametrically, $\lambda$, dependent potential or a dissipative field, ${\bf F}_e$, in the same way that we generated probability distributions in section 2 for $\Omega_t$. 

\begin{figure}
\includegraphics[scale=0.8]{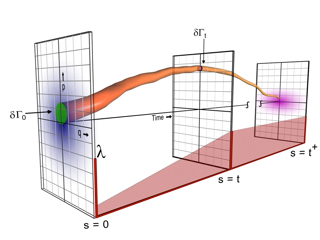}
\caption{
An illustration of a set of deterministic trajectories generated under the action of an external agent, where the agent does work on the system of magnitude $W\pm dW$.  The external agent is represented as a time-dependent $\lambda$ parameter which controls the system's potential energy and equilibrium state.  The trajectories are initiated under equilibrium conditions under a constant $\lambda(s=0)$, say equal to $A$; the shading in the $s=0$ phase-space plane being representative of the equilibrium distribution, $f_{eq}({\bf \Gamma}, \lambda=A).$  Here we represent a  linearly time-dependent $\lambda$ which varies over the time interval $0 \le s \le t$, with $W=\int_0^t ds \dot{\lambda}\partial{\cal H}/\partial \lambda$.   We've illustrated a case where $\lambda$ is not applied quasi-statically; {\it i.e.}, at time $t$ when the field has attained its final value of $\lambda=B,$ the system is far-from-equilibrium, and relaxes over $t \le s \le t^+$ towards to a new equilibrium state characterised by $\lambda=B$.  The magnitude of $f_{eq}({\bf \Gamma}, \lambda=B),$  the equilibrium distribution for $\lambda=B$, is represented by the shading in the $s=t^+$ plane.
The change in free energy brought about by $\lambda=A \rightarrow \lambda=B$ is  determined by the ratio of the integrals over ${\bf \Gamma}$ of $f_{eq}$ at $\lambda=B$ and $\lambda=A$.
}
\label{fig:crookfw}
\end{figure}

\begin{figure}
\includegraphics[scale=0.7]{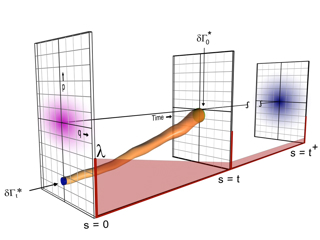}
\caption{An illustration of a set of time-reversed trajectories, conjugate to trajectory segments in the  time interval $0 \le s \le t$ in (a).   These trajectories are constructed under a time-reverse mapping as explained in Figure 1, where the field is also time-reversed, $\lambda(s=0)=B \rightarrow \lambda(s=t)= A$.  As these are generated under deterministic dynamics, this bundle of reverse trajectories will be characterised by the same magnitude, but opposite sign of $W$ as in (a), {\it i.e.}, $-W\mp W$.  Notice, that the initial coordinates of these reverse coordinates are not significantly weighted in $Z_B$, the partition function of the equilibrium state at $\lambda=B$; {\it i.e} $f_{eq}({\bf \Gamma}_t^\ast, \lambda=B)$ is small and these reverse trajectories, sampled from an initial equilibrium state,  are ``rare".  These reverse trajectories become less rare as the difference in A and B becomes small, and $\dot \lambda (s)$  over the interval $0 \le s \le t$ is reduced.  This provides a practical challenge when sampling the distribution $p_r(W=-{\cal A})$ in Crooks FT, or ensuring convergence in the average $\langle \exp{[-\beta W]} \rangle_f$ in Jarzynksi's equality.   However, it does not negate the validity of either relation.  One may also note that this same sampling problem can occur with Evans-Searles FT also.  However, in Evans-Searles FT, the external agents must have even time parity, it is the ``fast" application of external agents that render rare trajectories with $\Omega_t < 0$ that can be difficult sample in experiment or simulation.}
\label{fig:crookrv}
\end{figure}

In contrast to the Evans-Searles FT, the Crooks FT considers the probability of observing trajectories from two different equilibrium states. The probability density for a trajectory of duration $t$, initiated at equilibrium with $\lambda =A$, is
\begin{equation} 
p_f(W={\cal A}) = \int_{D_A} d{\bf \Gamma}_0 \, \delta\big[ W({\bf \Gamma}_0, t)-{\cal A}\big] f_{eq}({\bf \Gamma}_0, \lambda=A),
\end{equation}
where $W({\bf \Gamma}_0, t)$, denotes the work done over a trajectory of duration $t$, initiated at ${\bf\Gamma}_0$, and $f_{eq}({\bf \Gamma}_0, \lambda=A)$ is the equilibrium distribution in the canonical ensemble.
Now the reverse trajectory or anti-trajectory starts at coordinates ${\bf \Gamma}^\ast_t$, and is guaranteed under deterministic dynamics to give a value of work that is equal and opposite that of the forward trajectory, see Figures \ref{fig:crookfw} and \ref{fig:crookrv}. In this case, the reverse trajectory must also initiate under equilibrium conditions, however with $\lambda=B$, and the time-dependence of the parameter $\lambda$ and field ${\bf F}_e$ must be reversed compared to the forward trajectory.

So, the probability density for this reverse trajectory is
\begin{equation}
p_r(W= -{\cal A})  =  \int_{D_B} d{\bf \Gamma}^\ast_t \, \delta\big[ W({\bf \Gamma}^\ast_t, t)+{\cal A}\big] f_{eq}({\bf \Gamma}^\ast_t, \lambda=B). 
\end{equation}
At this point is it useful to note that if the system is driven strongly, {\it i.e.} far-from-equilibrium, the destination coordinate in the forward trajectory, ${\bf \Gamma}_t,$ may not be significantly weighted in the equilibrium distribution associated with the initial coordinates of the reverse trajectory; or in other words,  
\begin{equation}
f_{eq}({\bf \Gamma}_t, \lambda=B) =  \frac{ \exp{\big[-\beta{\cal H}_{\lambda=B}({\bf \Gamma}_t) \big ]}} {Z_B} 
\end{equation}
may be very small.
That is, the reverse trajectory can be ``rare", creating a difficult challenge in sampling the distribution $p_r(W=-{\cal A})$ in the Crooks FT, causing the convergence of the ensemble average in the Jarzynski Equality to become very slow \cite{Jarzynski-PRE-2002,Kofke-FPE-2005, Kofke-MP-2006,Kosztin-JCP-2006,Jarzynski-PRE-2006}. Recasting  ${\cal H}_{\lambda=B}$ in terms of the work done on the system, using the first law, {\it i.e.,}
\begin{eqnarray*}
W( {\bf \Gamma}^\ast_t,t) & = & -W( {\bf \Gamma}_0,t) 
= -\int_0^t ds \dot{\cal H}^{ad}({\bf \Gamma}_{s}, \lambda(s)) \\ 
 & = &  \bigg[{\cal H}_ {\lambda=A}({\bf \Gamma}_0) - {\cal H}_{\lambda=B}({\bf \Gamma}_t)\bigg] + k_BT\int_0^t ds \Lambda(s),
\end{eqnarray*}
and noting that $\cal H$ is an even function of the momenta, {\it i.e.} ${\cal H}_{\lambda}({\bf \Gamma}^\ast_s) = {\cal H}_{\lambda}({\bf \Gamma}_s)$ we see,  
\begin{eqnarray*}
f_{eq}({\bf \Gamma}^\ast_t, \lambda = B) 
 =  \frac{Z_A}{Z_B} f_{eq}({\bf \Gamma}_0,\lambda=A)\exp{\bigg[\beta W({\bf \Gamma}^\ast_t, t)\bigg]}\frac{\delta V({\bf \Gamma}_0)}{\delta V({\bf \Gamma}^\ast_t)} \end{eqnarray*}
where we have used the phase space compression factor given earlier. 
Now $\delta V({\bf \Gamma}_s^\ast)=\delta V({\bf \Gamma}_s)$ and the Jacobian is such that, 
\begin{equation}
p_r(W=-{\cal A})  = \frac{Z_A}{Z_B} \int_{D_A} d{\bf \Gamma}_0 \, \delta\big[ W({\bf \Gamma}^\ast_t, t)+{\cal A}\big]    \exp{\bigg[\beta W({\bf \Gamma}^\ast_t, t)\bigg]}   f_{eq}({\bf \Gamma}_t, \lambda=A). 
\end{equation}
Furthermore, as forward and reverse trajectories have equal but opposite values of $W$ under time-reversible dynamics, $W({\bf \Gamma}^\ast_t, t) =  -W({\bf \Gamma}_0, t),$ thus
\begin{eqnarray*}
p_r(W=-{\cal A})  =  \exp{\bigg[-\beta (\Delta F- {\cal A})\bigg]} \int_{D_A} d{\bf \Gamma}_t \, \delta\big[ W({\bf \Gamma}_t, t)-{\cal A}\big]   f_{eq}({\bf \Gamma}_t, \lambda=A).
\end{eqnarray*}
The integral on the RHS can be identified as $p_f(W={\cal A})$, resulting in the Crooks FT:
\begin{equation}
\frac{p_f(W={\cal A})}{p_r(W=-{\cal A})}= \exp{\bigg[\beta (\Delta F- {\cal A})\bigg]}.
\end{equation}
In the above derivation of the Crooks FT for the deterministic system it was assumed that :
\begin{enumerate}
\item{the dynamics is such that any phase point ${\bf \Gamma}$ for which $f({\bf \Gamma},\lambda=A)\ne 0$, $f({\bf \Gamma}_t^\ast,\lambda=B)\ne 0$ } 
\item{$f({\bf \Gamma},0) = f({\bf \Gamma}^\ast,0)$, and}
\item{when the time evolution of $\lambda$ and ${\bf F}_e$ are reversed, the dynamics remain deterministic and microscopically time-reversible.}
\end{enumerate}
These are sufficient conditions for the Crooks FT to be valid, but the condition of reversibility can be relaxed to some degree, and stochastic versions this FT exist.

\subsection{Steady-State Fluctuation Theorems}

Thus far, we have focused upon FTs that apply to a system driven out of an initial equilibrium state by an external field, characterised by ${\bf F}_e$, or a parametric change in the potential characterised by $\dot\lambda \ne 0$.  Indeed, the Evans-Searles FT applied to systems driven from a known initial state over transient trajectories,  is often referred to as the Transient Fluctuation Theorem (TFT).
However, according to the derivations of the Evans-Searles FT, the initial phase-space distribution is not restricted to time-invariant or even equilibrium distributions.  The only requirement the Evans-Searles FT places on the initial distribution function is that it is known and expressible in the ostensible dimension of the equations of motion (This is not the case for Crooks FT).  Here we consider the FTs applied to trajectories under a steady-state; {\it i.e.}, the system is acted upon by a purely dissipative, constant external field, ${\bf F}_e$.

There are two Steady-Sate Fluctuation Theorems (SSFTs)  that appear in the literature.  Both can be traced back to the original paper on FTs \cite{Evans-PRL-1993,Evans-PRL-1993a} that focused upon isoenergetic equations of motion; but it is only later  that two separate theorems were distinguished: (i) the steady-state version of the Evans-Searles FT \cite{Evans-AP-2002}  and (ii) the Gallavotti-Cohen FT \cite{Gallavotti-PRL-1995}.

\subsubsection{The Evans-Searles Steady State Fluctuation Theorem (SSFT)}

In its simplest formulation, the  SSFT of Evans \& Searles involves a rearranged from of eqn~\ref{eqn:FT} applied in the long time limit to trajectories of a system wholly in a nonequilibrium steady-state; {\it i.e.} the distribution function is time-invariant.   A more complete derivation of the SSFT valid under conditions including the decay of correlations is available
\cite{Searles-JSP-2007}; however here we provide a simpler presentation that is  physically compelling, and suitable  
for those primarily interested in a scientific 
justification. 

The argument of the Evans-Searles FT, applied to the steady state is
\begin{eqnarray}
\Omega_t^{ss}({\bf \Gamma}_0) &=& \ln{\bigg[\frac{dP(d{\bf \Gamma}_0,0)}{dP(d{\bf \Gamma}^\ast_t,0)}\bigg]}\nonumber\\
& =&  \ln{\bigg[\frac{f^{ss}({\bf \Gamma}_0,0)}{f^{ss}({\bf \Gamma}^\ast_t,0)}\bigg]}-\int_0^t\Lambda({\bf \Gamma}_s,s)ds
\label{eqn:omegass}
\end{eqnarray}
where $f^{ss}(d{\bf \Gamma}_0,0)$ is now the phase-space distribution function  associated with a steady-state, rather than with an equilibrium state (as in the Evans-Searles FT applied to transient trajectories). However, typically this definition of $\Omega_t^{ss}$ is difficult to 
to implement. Firstly, steady-state distribution functions for the types of deterministic dynamics under consideration are not generally known.  What is known is that, in the steady-state, the dynamics approach a strange attractor that has a different fractal dimension to the ostensible phase space.  Even if we knew the details of this attractor, it would still be difficult to apply the Evans-Searles FT as it describes bundles of phase-space trajectories in the phase-space dimension and not in the dimension of the strange attractor. 
Note however, that there  are special cases where these steady-state distribution functions can be expressed simply and exactly under stochastic equations of motion, and eqn \ref{eqn:FT} may be applied \cite{Wang-PRE-2005}.  

In general these steady-state distribution functions are not known, and consequently, it is not possible to construct exact expressions for $\Omega_t^{ss}$ for deterministic trajectory segments of duration $t$ that are wholly at a nonequilibrium steady-state.  However an approximate steady-state dissipation function can be constructed from trajectories initiated at a known equilibrium, in the absence of the dissipative field, ${\bf F}_e$.  This distribution function is often referred to as the Kawasaki distribution function \cite{EvansMorriss}, and can be considered to form the basis of the formal proof\cite{Searles-JSP-2007}.  At time $t=0$, the dissipative field is introduced, and we can express $\Omega_t$ associated with this trajectory in terms of its instantaneous rate of change, $\Omega(s)$ at time $s$:
\begin{equation}
\Omega_t = \int_0^\tau ds \Omega(s) + \int_\tau^t ds\Omega(s).
\end{equation}
Here, $\tau$ is some arbitrary long time, say several Maxwell times, so that the fluid has completely relaxed into a steady-state.  Thus, $\Omega_t$ is cast as a sum of transient and steady-state contributions with  the steady-state contribution, identified as the steady-state dissipation function, $\Omega_t^{ss}$, used to approximate $\Omega_t$ with an error or order ${\cal O}({\tau})$ .  It is instructive to express these dissipation functions as time-averages, $\bar{\Omega}_t = \Omega_t/t$
such that 
\begin{equation}
\bar{\Omega}_t^{ss} \approx \bar{\Omega}_t + {\cal O}\bigg(\frac{\tau}{t}\bigg).
\end{equation}
We make the physically compelling argument that, in the long time limit, the distribution function for steady-state trajectories will asymptotically converge to that for the full transient trajectories:
\begin{equation}
\lim_{t\rightarrow \infty} p^{ss}(\bar{\Omega}_t^{ss})  = p(\bar{\Omega}_t). 
\end{equation}

However, the fluctuations in $\bar{\Omega}_t^{ss}$ also vanish in the long time limit, and, in order that the SSFT be of any importance, it is necessary that these fluctuations vanish more slowly than ${\cal O}(\tau/t)$, the error in the $\bar{\Omega}_t^{ss}$ approximation.  To argue that this is the case, we re-express $\bar{\Omega}_t^{ss}$ as a sum over contiguous trajectory segments of duration $\Delta t$:
\begin{eqnarray}
\bar{\Omega}_{t} & = & \frac{1}{t}\sum_{i=1}^{t/\Delta t}\int_{(i-1)\Delta t}^{i\Delta t}ds{\Omega}(s)\\
 & = & \frac{1}{t}\sum_{i=1}^{t/\Delta t}\Omega_{\Delta t}.\end{eqnarray}
 If $\Delta t$ is larger than the longest correlation time in the
system, then the sum $\sum\Omega_{\Delta t}$ is composed of independent
segments and the variance in the sum is proportional to the number
of segments or $t/\Delta t$. The factor $1/t$ in front of the sum
decreases the variance of the sum by a factor $t^{2}$. Thus, the
standard deviation of the steady
state dissipation function, $\bar{\Omega}^{ss}_{t}$, along a
steady-state portion of a trajectory is inversely proportional to
$\sqrt{t}$, and decays at a slower rate than that of the error in
the approximation of $\bar{\Omega}_{t}$ with $\bar{\Omega}^{ss}_{t}$.
Consequently, we can approximate $\bar{\Omega}_{t}$ in the FT, eqn \ref{eqn:FT},
with the steady-state dissipation function $\bar{\Omega}_{t}^{ss}$,
leading to the SSFT
\begin{equation}
\lim_{t\rightarrow\infty}\frac{1}{t} \ln \frac{p(\bar{\Omega}_{t}={\cal A})}{p(\bar{\Omega}_{t}=-{\cal A})}={\cal A}.
\label{eqn:SSFT}\end{equation}

\subsubsection{Gallavotti-Cohen FT}

The fluctuation relation of the Gallavotti-Cohen FT can be written 
\begin{equation}
\lim_{t\rightarrow\infty}\frac{1}{t} \ln \frac{p(-\bar{\Lambda}_{t}={\cal A})}{p({-\bar{\Lambda}}_{t}=-{\cal A})}={\cal A},
\label{eq:GCFT}\end{equation}
where the average phase space compression factor (or the divergence
of the flow), measured in the steady state, is given as\begin{eqnarray*}
\bar{\Lambda}_{t} & = & \frac{1}{t}\int_{0}^{t}ds\,\Lambda(s)\\
\Lambda(t) & = & \frac{\partial}{\partial\mathbf{\Gamma}}\cdot\dot{\mathbf{\Gamma}}.\end{eqnarray*}
The original  proposal of this FT \cite{Evans-PRL-1993} was made for the special case of isoenergetic dynamics, for which $\Omega({\bf \Gamma})=-\Lambda({\bf \Gamma})$.  However, subsequently
 the work of Gallavotti and Cohen \cite{Gallavotti-PRL-1995,Gallavotti-JSP-1995}
  strongly suggests that under appropriate conditions eqn~\ref{eq:GCFT} (and with a restriction on the values of $\cal A$ \cite{Gallavotti-MPEJ-1995}\footnote{Eqn \ref{eq:GCFT} is restricted to values of $A$ bounded by a value
${\cal A}^{\ast}$: ${\cal A}\in({-{\cal A}}^{\ast},{\cal A}^{\ast})$. In the small field limit this value is given to leading order as ${\cal A}^\ast=0+{\cal O}(F_e^2)$ \cite{Gallavotti-condmat-2004}.})
can be applied to a larger class of dynamical systems (e.g. constant
temperature systems). 
They arrived at eqn~ \ref{eq:GCFT}
through a formal derivation \cite{Gallavotti-PRL-1995,Gallavotti-JSP-1995}, which drew upon the 
Sinai-Ruelle-Bowen (SRB) measure (for a discussion see \cite{Young-JSP-2002}), which requires the dynamics to 
be an Anosov diffeomorphism \cite{Ruelle-2004}\footnote{An Axiom A diffeomorphism will also suffice for the Gallavotti-Cohen FT.}. A necessary but insufficient condition for this is that the dynamical
system must be hyperbolic \cite{Young-JSP-2002}. This means that
the number of expanding and contracting directions on the attractive
set must be equal, or in other words the number of positive and negative
finite time Lyapunov exponents must be equal and no zero exponents
are allowed. In general, the equations of motion,
eqn \ref{eqn:fulleom}, do not form an Anosov diffeomorphism.  To
address this Gallavotti and Cohen introduced a new hypothesis, termed
the chaotic hypothesis \cite{Gallavotti-1999}, \footnote{Quoting form \cite{Gallavotti-1999}, ``Chaotic hypothesis: for
the purpose of studying macroscopic properties, the time evolution
map $S$ of a many-particle system can be regarded as a mixing Anosov
map.'' In \cite{Gallavotti-PRL-1995,Gallavotti-JSP-1995} the term
``transitive Anosov map'' was used to mean ``mixing Anosov
map''.%
}, which, for the purposes of the  Gallavotti-Cohen FT, allows many-body dynamics to be treated as an Anosov diffeomorphism.
Unfortunately, as yet, there is no way to independently
ascertain if a physical system may be treated as Anosov diffeomorphic. The requirements for the valid application of the
Gallavotti-Cohen FT  to physical systems are therefore extremely difficult to establish.

There is a large body of computer simulation results, for various
processes, that have tested the steady state fluctuation theorems,
e.g. \cite{Evans-PRL-1993,Evans-PRL-1993a,Searles-PRE-1999,Ayton-JCP-2001,Evans-PRE-1995,Searles-JCP-2000,Searles-IJT-2001,Bonetto-PD-1997,Bonetto-C-1998,Bonetto-PRE-2001,Lepri-PRL-1997,Baranyai-JCP-2003,Schmick-PRE-2004,Dolowschiak-PRE-2005,Zamponi-JSP-2004,Williams-JCP-2006},
\footnote{There are results in the literature which test the Evans-Searles FT while erroneously
claiming to test the Gallavotti-Cohen FT.%
}. We known of no case for which the Gallavotti-Cohen FT converges faster than the
Evans-Searles FT. For temperature regulated dynamics, when the dissipative field
strength is very small, the Gallavotti-Cohen FT can take extremely long times to
converge. Indeed as the dissipative field strength approaches zero
the amount of time it takes the Gallavotti-Cohen FT to converge diverges \cite{Evans-PRE-2005,Gallavotti-condmat-2004}. To understand
this consider the arguments of the Evans-Searles FT and the Gallavotti-Cohen FT. When the field
strength approaches zero so does the instantaneous dissipation function.
More precisely the average value of the instantaneous dissipation
function, to leading order, is $\Omega(\mathbf{\Gamma})=0+\mathcal{O}(F_{e}^{2})$,
and the standard deviation is $\sigma=0+\mathcal{O}(F_{e})$. For
the phase space compression factor, the mean is $\Lambda=0+\mathcal{O}(F_{e}^{2})$
and the standard deviation is $\sigma=\sigma_{0}+\mathcal{O}(F_{e}^{2})$,
where $\sigma_{0}$ is the amplitude of the standard deviation at
equilibrium. The difference between the behaviour in the amplitude
of the fluctuations, $\sigma$, for these two quantities is crucial.
As $F_{e}$ approaches zero so to does the amplitude of the fluctuations
in $\Omega$ but not those in $\Lambda$. Now the form of the fluctuation
formulae is asymmetric. In the limit $F_{e}\rightarrow0$, eqn \ref{eqn:SSFT}
(the SSFT) remains consistent with a given trajectory segment being
equally likely to occur as its anti-trajectory segment. This is a
necessary condition for equilibrium. In contrast eqn \ref{eq:GCFT}
(the Gallavotti-Cohen FT) is not consistent with this, due to the fluctuations in
$\Lambda$ remaining finite when $F_{e}\rightarrow0$. One way
this could be resolved is for the time averaging, or the time for which
the Gallavotti-Cohen FT is given to converge, to be so long that there are no significant
fluctuations remaining. However it is specified by the theory \cite{Gallavotti-condmat-2004} that the largest fluctuations for which the Gallavotti-Cohen FT may be validly applied vanishes in the small field limit.

\subsection{The Einstein Relation and Green Kubo Theory}

The Evans-Searles Fluctuation Theorem, as well as Crooks Fluctuation Theorem have been reviewed here as recent theorems in nonequilibrium statistical mechanics.  Here we show that the FTs, and in particular the Evans-Searles FT, is completely consistent with the long-standing and well-known relations in the field, namely  the Einstein-Sutherland relation \cite{Einstein-AP-1905,Sutherland-PM-1905, Einstein-1956} and the Green-Kubo Relations.
The Einstein-Sutherland relation dates back to the very early days of non-equilibrium statistical 
mechanics; it can be written as
\begin{equation}
\left\langle \mathbf{v}\right\rangle _{F_{e}}=\beta D\mathbf{F}_{e}.\label{eq:GK-SuthE}
\end{equation}
This important relation describes  the average steady-state velocity of a particle, $\left\langle \mathbf{v}\right\rangle _{F_{e}}$, under an applied field, ${\bf F}_e$, to the variance in the particle's displacement over time in the absence of the field, which is commonly referred to as  the diffusion constant, $D=\lim_{t\rightarrow\infty}\left\langle \Delta x(t)^{2}\right\rangle _{0}/(2t)$, where $\Delta x(t)=\int_{0}^{t}ds\;\dot{x}(s)$. 
 Starting from the Evans-Searles FT, we reformulate a generalised form of this Einstein-Sutherland relation, and from that, the Green-Kubo relations.  
While this does not produce new results, it demonstrates the FTs' consistency with important existing theorems in nonequilibrium statistical mechanics, and it also emphasises/clarifies the conditions necessary for the application of these theories, as we show later, in the case of supercooled liquids. 

To derive the more generalised version of the Einstein-Sutherland relation, eqn \ref{eq:GK-SuthE}, from the FT, we first need to identify the product of the particles' drift velocity and the applied field, $-{\bf F}_e\cdot {\bf v}(t)$, as a specific example of a  dissipative field flux, that we have represented by ${\bf J}V\cdot{\bf F}_e$, which, in the case of the flux and constant field being in the same direction, we write more simply as $J V F_e$.   Under steady-state, the time-averaged dissipative flux is defined as 
\begin{equation}
\bar{J}_{t}\equiv\frac{1}{t}\int_{\tau}^{\tau+t}ds\; J(s),\label{eq:GK-flux}\end{equation}
 where $\tau$ is a time long enough after the application of the field so that 
the system is at a steady-state. The steady-state
fluctuation theorem, eqn \ref{eqn:SSFT}, may then be written as
\begin{equation}
\lim_{t\rightarrow\infty}\frac{1}{t}\ln\frac{P(\bar{J}_{t}={\cal A})}{P(\bar{J}_{t}=-{\cal A})}=-\beta VF_{e}{\cal A}.\label{eq:GK-SSFT}
\end{equation}
In the limit of long time, we may invoke the central limit theorem, which
states that close to the mean the distribution of ${\bar J}_t $ will be Gaussian. Additionally in the limit of small field strength, values of ${\bar J}_t$ close to the mean will dominate:
\begin{equation}
P\left(\bar{J}_{t}={\cal A}\right)=\frac{1}{\sigma\sqrt{2\pi}}\exp\left(\frac{\left({\cal A}-\left\langle {\bar J}_t\right\rangle _{F_{e}}\right)^{2}}{2\sigma^{2}}\right),\label{eq:GK-Gauss}\end{equation}
Now as the variance in the distribution of ${\bar J}_t$ is independent of the field direction or sign of  $F_e$, then, to leading order in ${F}_e$, the variance behaves as
\begin{equation}
\sigma_{F_{e}}^{2} =  \sigma_0^2+\mathcal{O}\left(F_{e}^{2}\right) 
 =  \left\langle \bar{J}_{t}^{2}\right\rangle _{0}+\mathcal{O}\left(F_{e}^{2}\right).\label{eq:GK-variance}
\end{equation}
The SSFT, eqn \ref{eq:GK-SSFT}, the central limit theorem eqn \ref{eq:GK-Gauss} and eqn \ref{eq:GK-variance}
combine to give a generalised Einstein-Sutherland relation:
\begin{equation}
\left\langle J\right\rangle _{F_{e}}=\lim_{t\rightarrow\infty}-\frac{1}{2}\beta VF_{e}\left\langle \bar{J}_{t}^{2}\right\rangle _{0}t+\mathcal{O}(F_{e}^{2}),\label{eq:GK-gen-Einstein}
\end{equation}

\begin{figure}
\includegraphics[bb=0bp 0bp 828bp 556bp,clip,scale=0.8]{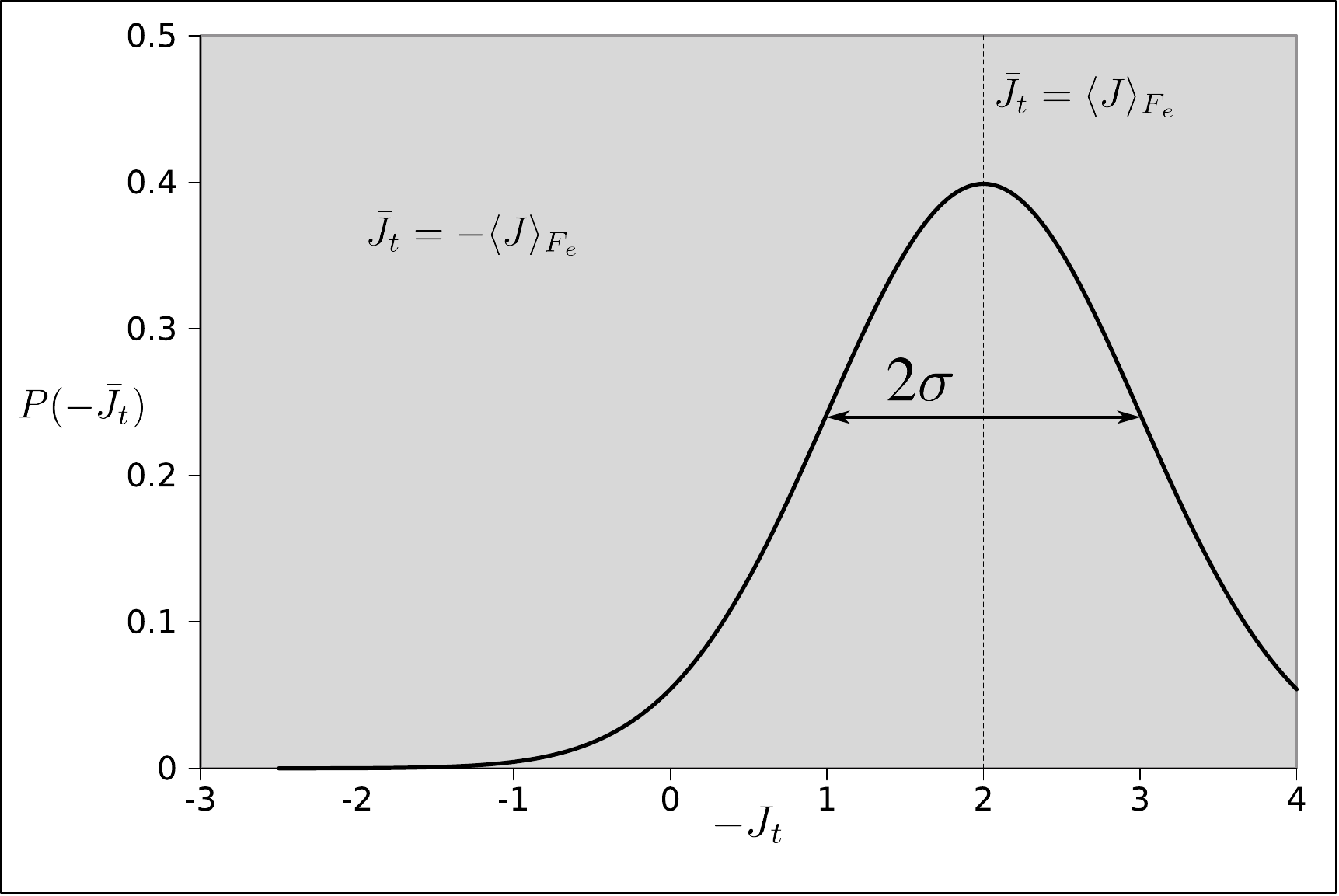}
\caption{A typical Gaussian distribution, of unit variance $\sigma^{2}=1$,
for the time averaged flux $\bar{J}_{t}$ with the ensemble average
$\langle J\rangle_{F_{e}}=-2$ denoted by the dashed line. The other
dashed line is at $\bar{J}_{t}=-\langle J\rangle_{F_{e}}=2$ which
is the value the fluctuation theorem compares to the mean. As time
proceeds, in the limit $t\rightarrow\infty$, the variance decreases
like $\sim1/t$.\label{fig:Gaussian}}
\end{figure}

It is generally known that the Einstein-Sutherland relation is only valid to linear order in the field, $F_e$. However, the FT provides more detailed understanding of how this relation fails under large fields.  When the field is increased, the mean dissipative flux, $\langle {\bar J}_t \rangle$ will also increase,  Figure~\ref{fig:Gaussian}. When the mean is large
relative to the standard deviation, then for every typical value of the flux $\bar{J}_{t}={\cal A}$, its conjugate value $\bar{J}_{t}=-{\cal A}$ in the SSFT will be represented in the wings of the distribution where the central limit theorem  no longer applies.  In this instance, the generalised Einstein-Sutherland relation is invalid, even if the equilibrium variance $\langle {\bar J}^2_t \rangle_0$  is replaced with the variance under steady-state, or $\langle {\bar J}_t^2\rangle_{F_{e}}$.
Molecular
dynamics simulations of planar shear have shown that the breakdown of the central limit theorem dominates over the approximation of ignoring terms of ${\cal O}(F_e^2)$ or higher in the variance that describes the distribution near its mean. \cite{Searles-JCP-2000}.
In the case of a single, tagged particle, interacting with a constant field,
and embedded in a supercooled liquid, the amount of time it takes for the the steady-state FT 
to converge as well as the the time it takes for the distribution to become
Gaussian,  increases rapidly upon approach to the glass transition \cite{Williams-PRL-2006}.
Moreover, the variance decreases with time.  
As this time increases the variance decreases inversely proportionally.
As the nominal glass transition is approached the strongest field
for which a linear response may be observed in the steady state,
vanishes. 

The variance of the flux may be expressed in terms of the integral,\
\begin{eqnarray}
\left\langle \bar{J}_{t}^{2}\right\rangle _{0} & = & \frac{1}{t^{2}}\int_{0}^{t}ds\int_{0}^{t}du\;\left\langle J(s)J(u)\right\rangle _{0}\\
 & = & \frac{2}{t}\int_{0}^{t}ds\;\left\langle J(0)J(s)\right\rangle _{0}-\frac{2}{t^{2}}\int_{0}^{t}ds\;{s\left\langle J(0)J(s)\right\rangle }_{0}.\end{eqnarray}
 In the long time limit, the second term on the RHS vanishes and
 \begin{equation}
\lim_{t\rightarrow\infty}\left\langle \bar{J}_{t}^{2}\right\rangle _{0}=\frac{2}{t}\int_{0}^{t}ds\;\left\langle J(0)J(s)\right\rangle _{0}.\label{eq:GK-GK}\end{equation}
 Combing this with the generalised Einstein-Sutherland relation, eqn \ref{eq:GK-gen-Einstein}, gives
 the celebrated Green-Kubo theory for steady-state:
 \begin{equation}
\left\langle J\right\rangle _{F_{e}}=-\beta VF_{e}\lim_{t\rightarrow\infty}\int_{0}^{t}ds\;\left\langle J(0)J(s)\right\rangle _{0}
\end{equation}
It can be used to obtain a transport coefficient in
terms of equilibrium fluctuations in the form of an autocorrelation
function. One might wonder why the Green-Kubo theory holds such important
status given that the presentation here shows it to be equivalent
to the Einstein-Sutherland relation. In contrast to the Einstein-Sutherland
relation the Green-Kubo theory is also applicable to time dependent
phenomena. A time dependent version of the Green-Kubo theory cannot
be obtained from the FT which must satisfy definite time parity conditions.

An example of where the FT has been used is Couette flow or planar shear \cite{Evans-PRL-1993, Searles-JCP-2000} using the SLLOD equations of motion\cite{EvansMorriss}, which in the absence of a fictitious
thermostat and for the case of constant
shear rate $\dot{\gamma}$, are equivalent to Newtons equations of
motion. As we are controlling the shear rate externally we identify it as the external
field $F_{e}=\dot{\gamma}$ and the flux as $J=P_{xy}$, where $P_{xy}$ is the $xy$ element of the pressure tensor. The dissipative field or entropy production for Couette flow is then
\begin{equation}
JVF_{e}=\dot{\gamma}VP_{xy}
\end{equation}
The shear viscosity $\eta$ is the rate at which work is being done on the fluid
divided by the product of the volume and the shear rate squared.
The Green Kubo expression for the viscosity is thus \begin{equation}
\eta=\frac{-\left\langle P_{xy}\right\rangle }{\dot{\gamma}}=\beta V\int_{0}^{\infty}dt\;\left\langle P_{xy}(0)P_{xy}(t)\right\rangle _{0},\label{eq:GK-viscosity-GK}\end{equation}
and the Einstein-Sutherland expression is
\begin{equation}
\eta=\frac{1}{2}\beta Vt\lim_{t\rightarrow\infty}\left\langle \bar{P}_{xy,t}^{2}\right\rangle .\label{eq:GK-viscosity-Einst}\end{equation}
If the system is very viscous, the Green-Kubo expression, eqn \ref{eq:GK-viscosity-GK}, will require a long time to converge and consequently, the generalised Einstein-Sutherland expression, eqn \ref{eq:GK-viscosity-Einst}, will
probably be the better method to extract the viscosity in such a situation \cite{Hess-PRE-2001}.
In contrast, if we wish to calculate the self-diffusion coefficient
for a very viscous system, the Einstein-Sutherland expression for the diffusion $\left\langle \Delta{x(t)}^{2}\right\rangle _{0}/(2t)$,
can be very slow to converge while the Green Kubo expression,
$D=\int_{0}^{\infty}ds\left\langle v(0)v(s)\right\rangle _{0}$,
 will usually converge quite rapidly \cite{Williams-PRE-2006}.

\section{Applications of Fluctuation Theorems to Experimental \& Model Systems}

Much of the work done in developing and extending the Fluctuation Theorems was accomplished by theoreticians and mathematicians interested in non-equilibrium statistical mechanics.  Until 2002, demonstrations of the theorems were limited to computer simulations and there were no practical experimental demonstrations of the theorems, despite the range of interests in nano/micro machines, or molecular devices that impose nanometer scale displacements with picoNetwon scale forces.   Such small machines include single biomolecules that act as molecular motors  and whose experimental observation highlight the nonequilibrium phenomena described by the FTs.   Linear motors, such as the action-myosin or the kinesin-microtubule motor are fuelled by proton currents or ATP hydrolysis and  function as integral parts of cellular metabolism, and consequently, they work under inherently nonequilibrium conditions\cite{Howard-2001}.    Over time, on average, these molecular engines must not violate the Second Law; however occasionally they run ``backwards", converting heat from the surroundings to generate useful mechanical/chemical energy.  This work, done on the molecular time and length scales, will have a natural variation or spread of values, and the conjecture is that this 
is governed by the FTs.

In 2002 the FTs were demonstrated experimentally by two independent groups, each with a unique focus and both using optical tweezers.  Wang {\it et al} \cite{Wang-PRL-2002} demonstrated the Evans-Searles Fluctuation Theorem by monitoring the transient trajectory of a single colloidal bead in a translating optical trap.  Simultaneously, Liphardt {\it et al}\cite{Liphardt-S-2002}, used optical tweezers to pull the ends of a DNA-RNA  hybrid chain, measuring the work required to unravel or unfold a specific domain in the chain.  These experiments had complementary aims:  The colloidal experiment was a classical model system constructed to cleanly demonstrate, as rigorously as possible in experiment, the Evans-Searles FT.  In contrast Liphardt's  RNA-unfolding experiment importantly demonstrated the application of Crooks FT to a complex biomolecular system, highlighting the potential practical use of FTs to a wider range of scientists.  In this section we review both experiments in some detail before more briefly mentioning other more recent experimental applications of the FTs, as well as other proposed experimental systems and implications.

\subsection{Single Colloidal Particle in an optical trap}

An optical trap is formed when a transparent, micron-sized particle, whose index of refraction is greater than that of the surrounding medium, is located within a focused laser beam.  The refracted rays differ in intensity over the volume of the sphere and exert a subpico-Newton force on the particle, drawing it towards the region of highest intensity, {\it i.e.}, the focal point or trap center.  The optical trap is harmonic; a particle located a distance ${\bf r}$ from the center of the trap has an optical force, ${\bf f}_{opt}=-k{\bf r}$, acting to restore its position to the trap center.  $k$ is the trapping constant which is determined by the distribution of particle positions at equilibrium and is tuned by adjusting the intensity of the laser.   Using an objective lens of high numerical aperture, the optical trapping is strongest in the direction perpendicular to the focal plane, such that particle remains localised entirely within the focal plane, fluctuating about the focal point.    As the particle position ${\bf r}$ is measured at mHz frequency, {\it i.e.} over timescales significantly large that inertia of the colloidal particle  is negligible,  the measured optical force, ${\bf f}_{opt},$ balances any applied forces, either forces arising from the surrounding solvent, such as Brownian or drag forces, or the tension associated with a tethered chain molecule such as DNA or RNA.

The first experiment that demonstrated the Fluctuation Theorems was carried out by Wang {\it et al} \cite{Wang-PRL-2002}.  They monitored the trajectory of a single colloidal particle,  weakly held in a stationary optical trap that was translated uniformly with constant, vanishingly small velocity ${\bf v}_{opt}$ starting at time  $s=0$.  Initially, the particle's position in the trap is distributed according to an equilibrium Boltzmann distribution with an average particle velocity of 0.  With trap translation, the particle is displaced from its equilibrium position until, at some time later, the average velocity of the particle is equal to the velocity of trap.  From this point the system is in a nonequilibrium steady state.  To determine the dissipation function, consider that the the external field is purely dissipative, {\it i.e. } $\dot \lambda = 0$ and ${\bf F}_e \sim {\bf v}_{opt}$, so that the
dissipation function is
\begin{equation}
\Omega_t = -\beta \int_0^t ds {\bf J}V\cdot{\bf F}_e = \beta
\int_0^t ds {\bf f}_{opt}\cdot {\bf v}_{opt}.
\end{equation}
With the ability to resolve nanometer-scale particle displacements and femtoNewton scale optical forces, $\Omega(s)$ was determined with sub-$k_BT$ resolution.   As there is no change in state of the underlying state of the system, and the field has even time-parity, the Evans-Searles FT  and the Crooks FT reduce to the same expression and equivalently describe the distributions of $\Omega_t \equiv W.$  We expect, from the Second Law, that work is done to translate the particle-filled optical trap, or $W>0$, but according to the FTs there should also be a nonvanishing probability of observing short trajectories where $W< 0$, that is thermal fluctuations {\it provide} the work.   Indeed, Wang {\it et al} showed trajectories  with $W = \Omega_t < 0$, persisted for 2-3 seconds, far longer than had been demonstrated by simulation.  However, in this initial 2002 experiment  there was an insufficient number of  trajectories  to properly sample the full distribution, $p(\Omega_t)$, and the authors instead tested a coarse-grained form of the Evans-Searles FT,  the integrated Evans-Searles FT:
\begin{equation}
\frac{p(\Omega_t < 0)}{p(\Omega_t > 0)} = \langle \exp[{-\Omega_t}] \rangle_{\Omega_t>0},
\end{equation}
where the brackets on the RHS denote the average over that part of the distribution for which $\Omega_t > 0$.  Later, Wang et al \cite{Wang-PRE-2005} revised this same experiment, and sampled a larger number of trajectories, enabling a direct demonstration of the Evans-Searles FT with a purely dissipative field.  Moreover, they also translated the particle-filled optical trap in a circular or ``race-course" pattern, producing one long single trajectory, which outside of the initial short time interval, was steady-state.  Using contiguous segments of this single steady-state trajectory, they demonstrated the steady-state version of the Evans-Searles FT \cite{Wang-PRE-2005}.

A particle in an optical trap was also used to demonstrate the distinction between the Evans-Searles FT and the Crooks FT, in the so called ``capture" experiment \cite{Carberry-PRL-2004, Reid-PRE-2004}.   In this single particle experiment, the strength of the stationary optical trap, or the trapping constant $k$,  is changed instantaneously, and the time-dependent relaxation of the particle position from one equilibrium distribution to another distribution is recorded.  For this experiment, a particle is localised in a stationary trap of strength $k_0$ over a sufficiently long time that its position is described by an equilibrium distribution.  At time $s=0$, the optical strength is increased discontinuously from $k_0$ to $k_1$, $k_1 > k_0$, so that we more tightly confine or ``capture" the particle.  Alternatively, we can decrease the trap strength from $k_1$ to $k_0$, to ``release" the particle.  Thus, the external field parameter, $\lambda,$ is the time-dependent trap strength, $k(s)$, which varies discontinuously, $\dot{\lambda}(s) = (k_1-k_0)\delta(s),$ and in the absence of a purely dissipative field, or ${\bf F}_e={\bf 0}.$  The particle's position is recorded as it relaxes to its new equilibrium distribution and the different functions $W$ and $\Omega_t$ are evaluated over an ensemble of nonequilibrium trajectories.

Work is the change in the internal energy that occurs with the change in the trapping constant: $$W = \int_0^t ds {\dot\lambda}\frac{d\phi_{ext}}{d\lambda} = 1/2(k_1-k_0){\bf r}_0^2.$$  Note that $W$ will always be positive if the trap strength is $k_1>k_0$ and consequently, distributions for $W$ cannot be Gaussian.  As all trajectories must initiate under equilibrium conditions under $k_0$, the probability distribution of ${\bf r}_0$ is a Boltzmann distribution and the distribution of $W$ is then simply
$$p_{k_0\rightarrow k_1}(W)= \sqrt{\frac{k_0}{\pi(k_1-k_0)W} }\exp{\bigg[-\frac{k_0W}{k_1-k_0}\bigg]}.$$
Thus, if we consider the ensemble average, $\langle \exp{[-\beta W]} \rangle$, we would recover the Jarzynski Equality, or $\langle \exp{[-\beta W]} \rangle = \exp{[-\beta \Delta F]},$ where $\Delta F = k_BT \ln{[\sqrt{k_1/k_0}]}$ from classical thermodynamics.    Furthermore, if we consider the probability distribution of $W$ for both the {\it forward} or capture direction, and the {\it reverse} or release direction, {\it i.e.}, $W_{k_1\rightarrow k_0}$,
$$p_{k_1\rightarrow k_0}(W)= \sqrt{\frac{k_1}{\pi(k_0-k_1)W} }\exp{\bigg[-\frac{k_1W}{k_0-k_1}\bigg]},$$
it is straightforward to show that these distributions trivially obey Crooks FT.    Notice that in the context of the capture experiment, Crooks FT depends only upon the equilibrium distribution of initial particle distribution: these equilibrium distributions are independent of, for example, the viscoelastic response of the suspending fluid.  An alternative experiment, say where the trapping constant changes linearly over some time period is different; there $W$ is accumulated over the time period over which $k$ is changing and the distribution of $W$ depends upon the response or microrheology of the fluid.

In contrast, the dissipation function, $\Omega_t$ depends sensitively upon the non-equilibrium trajectory,  and upon the material properties of the surrounding fluid \cite{Carberry-JOPA-2007}.  The dissipation function is then
$$\Omega_t= \int_0^t ds {\dot{\bf r}}\bigg[\frac{d\phi_{ext}(\lambda=k_0)}{d{\bf r}} - \frac{d\phi_{ext}(\lambda(s))}{d{\bf r}} \bigg] = \frac{k_0-k_1}{2}({\bf r}^2_t - {\bf r}^2_0).$$  Notice that if we consider a cyclic protocol where the external field had even time-parity, {\it i.e.} $\dot{\lambda}(s) = (k_1-k_0)\delta(s) - \delta(t-s)$, that is, we would ``capture" at $s=0$ and ``release" at $t=0$, then $\Omega_t = W_{k_0\rightarrow k_1} + W_{k_1\rightarrow k_0}$, and the Evans-Searles and Crooks  FTs would again reduce to the same expression.

These experiments were simple, but important demonstrations of the FTs.  These experiments can also be related to Fluctuation Theorems derived under stochastic equations motion, a topic not covered in this review.  The motion of a single colloidal particle in a purely viscous solvent is accurately described by a stochastic Langevin equation; {\it i.e.}, the degrees of freedom of the solvent molecules are integrated over the timescale of the colloidal particles'  motion to yield a friction coefficient and uncorrelated Gaussian noise.   As demonstrated by Wang {\it et al} \cite{Wang-PRE-2005}, from the stochastic, inertialess Langevin equation of motion,  you can construct the Evans-Searles FT in  much the same manner as was done for deterministic dynamics.  This is important evidence against the notion that  macroscopic irreversibility is due to the coarse-graining or separation of time-scales of the system's degrees of freedom: for the optically-trapped particles, both Langevin equations of motion as well as fictitious thermostatted deterministic equations (MD)  showed the monotonic decrease in $\Omega_t$ as the trajectories evolved in time.  On the other hand, if a system's dynamics are well described by stochastic Langevin dynamics, with uncorrelated Gaussian noise, then we are assured that the FTs will hold.  Indeed, several simple model experiments, including optically trapped colloidal particles in purely viscous media, can be described using stochastic Langevin dynamics  and consequently one could argue that these experiments confirm the Langevin dynamics rather than the FTs.  However more recently, Carberry{\it et al}, confirmed the FT using an optically trapped bead in a viscoelastic solvent, where the stochastic equations of motion require a dissipative term with memory, from which the FTs have yet to be directly derived.  This experimental confirmation is indeed a confirmation of the FTs, rather than a confirmation of the dynamics that satisfy the FTs.

\subsection{Stretching of Biopolymers}

In 2002 Liphardt {\it et al} \cite{Liphardt-S-2002} measured the tension-induced, unfolding transition of a P5abs domain in a single RNA molecule.  To do this they tethered the ends of a single DNA-RNA hybrid molecule containing the P5abc domain to micron-sized colloidal beads whose surfaces were chemically functionalised.  These beads act as "handles" to grab and manipulate the single molecule in an optical trap.  One bead is weakly held in an optical trap while the other bead is held in a micropipette whose position/translational speed is controlled by a piezoelectric actuator.  As the micropipette is translated relative to the optical trap, the chain is stretched and the tension in the chain is determined by the optical force, ${\bf f}_{opt}$ acting on the bead in the trap.  In this way, Liphardt {\it et al} constructed force-extension profiles of a single biomolecule, focussing specifically on a window of extensions over which the P5abs domain unfolded.  Over a $\sim30 \mu$m  extension range, they found that, typically, the force increased monotonically against the entropic elasticity of the chain's contour, but at $\sim 10 $pN the force was either constant or decreased slightly  for  a further $\sim10 \mu$m indicative of unfolding of the domain, before increasing monotonically again.  The time-reverse path corresponds to retraction of the chain ends, creating the same force-extension profile for the re-folding of the domain.   When the chain was unfolded slowly (corresponding to small $\dot\lambda$), the folding-unfolding processes were reversible, {\i.e.}, the work done (force integrated over extension) of the forward (unfolding) and reverse (re-folding) paths were roughly equal but of opposite sign.  When the chain was unfolded quickly, a hysteresis loop appeared in the forward-reverse force profile and the $W$ to fold and unfold differed in magnitude, due to the macroscopic irreversibility.  Theoretically, the distributions $p_f(W)$ and $p_r(W)$ at any given protocol or stretching rate, $\dot\lambda$, should obey Crooks FT: however, it is clear that variation in $W$ due to experimental error or approximations in analysis must be minimal in comparison to the inherent variation in $W$ that arises from the irreversibility of the process.   Furthermore, it is important to experimentally sample a sufficient number of trajectories, $N$: as the protocol rate, $\dot\lambda$, increases and the process is driven further away from a quasi-static or equilibrium process, the required number $N$ grows as ``rare" trajectories in the distribution become important as explained in Figure~\ref{fig:crookrv}.  Successive single molecule manipulations in an optical trap can be difficult, limiting the maximum possible number of trajectories.   However, Liphardt {\it et al} constructed ${\cal O}(300)$ stretching profiles, unfortunately an insufficient number to describe the distributions, but fortunately sufficient to  the show that  $\Delta F$ obtained from Jarzynski's equality could be determined to within $k_BT$ from their $N$ experimental trajectories.  In later work from the same group, Collin {\it et al}\cite{Collin-N-2005} similarly used optical tweezers to construct experimental work distributions and demonstrate Crooks FT for the folding/unfolding of an RNA hairpin and an RNA three-helix junction.  These results demonstrate that nonequilibrium single molecule measurements, when analysed in conjunction with the FTs can provide thermodynamic information, even though these single-molecule events may not be at equilibrium.   More importantly, these papers introduced to the single-molecule force spectroscopist, a proper analysis/interpretation of force measurements.

\subsection{Other model systems}

Several other more recent experimental demonstrations of the FTs have appeared in the literature in the last couple of years.  Garnier and Ciliberto \cite{Garnier-PRE-2005} demonstrated the FTs by measuring the fluctuating voltage of a resistor in parallel with a capacitor, driven out of equilibrium by a constant current flow;  Schuler {\it et al} \cite{Schuler-PRL-2005}  excited a single defect center in a diamond using an intensity-modulated  laser, forming a two-state system; and  Douarche {\it et al} \cite{Douarche-EL-2005, Douarche-PRL-2006} , experimentally checked the Jarzynski equality and Crooks FT against the thermal fluctuations of a mechanical oscillator in contact with a heat reservoir.  Applications of the FT to model systems need not be experimental.  To date the FT has been used in theoretical descriptions/computer simulations of sheared liquids {\it i.e.} \cite{Evans-PRL-1993}, chemical reactions \cite{Gaspard-JCP-2004, Seifert-JPA-2004, Andrieux-JCP-2004} molecular motors \cite{Seifert-EL-2005, Andrieux-PRE-2006}, granular gases \cite{Visco-JSP-2006}, and glasses \cite{Williams-PRL-2006, Williams-arXiv-2007}.

Finally, through its application to time-dependent shear of viscoelastic fluids, the FT has been used to resolve a long-standing problem in linear irreversible thermodynamics \cite{Williams-CRP-2007}.
The reference to $JVF_{e}$ as the entropy production is taken from
linear irreversible thermodynamics, which asserts that, in local equilibrium,
the entropy source strength is the sum of products of irreversible
thermodynamic fluxes and forces \cite{Kondepudi-Prigogine,DeGroot-Mazur}.
By the Second Law we then expect that $\left\langle J\right\rangle $
is always negative. However for time dependent processes in viscoelastic
fluids this is not always the case.  In the steady-state the free energy
$F$, the mean internal energy $U$, the temperature $T$ and the
entropy $S$ will all be constant.  In local equilibrium, where the
entropy transported to the reservoir is given by $\dot{S}_{tr}=\dot{Q}/T$,
we have $\dot{F}=-JVF_{e}+\dot{Q}-T\dot{S}=0$. This leads us directly
to the conclusion that $JVF_{e}=-T\dot{S}_{sp}$, where $\dot{S}_{sp}=\dot{S}-\dot{S}_{tr}$
is the rate at which entropy is being spontaneously produced, due
to the external field $F_{e}$. If we are not in the steady state
or are not in local equilibrium we cannot show that, $JVF_{e} = -T\dot{S}_{sp}$. Indeed for time dependent proccesses in viscoelastic fluids $\left< J \right>$ may be positive for periods of time.  
For a system initially in equilibrium, subject
to a purely dissipative process, {\it i.e.}, $\dot{\lambda}=0\;\forall\; t$,
the time averaged value of the entropy production,
rather than the more traditional instantaneous entropy production, can be shown from the Jarzynski Equality to form the inequality that specifies the time direction. This was demonstrated in simulations of  time-dependent planar shear of viscoelastic fluids \cite{Williams-CRP-2007} where the instantaneus value of $\left< J \right>$ was indeed shown to be negative for periods. Furthermore, this inequality
is valid for time dependent processes which may be arbitrarily far
from equilibrium.

\section{Conclusions}

This review has focused on two important fundamental fluctuation relations, the Evans-Searles and the Crooks FTs, with some discussion on steady state fluctuation relations (Evans-Searles SSFT and the Gallavotti-Cohen FT) and important relations that can be derived from these.  The new understanding of thermodynamic reversibility that has been obtained from the FTs has been described in detail, as has the determination of free energy differences between equilibrium states from study of nonequilibrium paths. In this context the Jarzynski equality has been presented.  

However, there is much work on FTs that has not been covered in this review.  The theoretical description above has focused on systems whose equilibrium equations of motion preserve a canonical distribution function.  This requirement is not necessary, and the application of the Evans-Searles FT to a wide range of combinations of ensembles and dynamics is well known \cite{Searles-JCP-2000a, Evans-AP-2002}. Furthermore, the Crooks FT has been extended to allow treatment of other ensembles including constant pressure ensembles \cite{Williams-MP-2007,Cuendet-JCP-2006, Chelli-JCP-2007}.  Granular systems \cite{Puglisi-JSM-2006} and systems undergoing chemical reactions \cite{  Andrieux-JSP-2007, Paramore-JCP-2007,  Schmiedl-JCP-2007,Andrieux-JCP-2004,Baranyai-JCP-2003} have also been studied.

As discussed above, we have focused on deterministic derivations due to our interest in emergence of irreversibility from reversible equations of motion, however there has been extensive work on stochastic versions of the Evans-Searles FT (see \cite{Lebowitz-JSP-1999, Searles-PRE-1999} for early work in this regard), and it is straightforward to see that the requirement of microscopic reversibility is not essential in its derivation - existence of the reverse trajectory is sufficient.  The original derivations of the Crooks FT and the Jarzynski equality used stochastic dynamics, and it is only more recently that derivations for deterministic systems have appeared \cite{Evans-MP-2003, Scholl-Paschinger-JCP-2006}.  As mentioned in the introduction, versions of FTs applicable to quantum systems have also been obtained \cite{Jarzynski-PRL-2004,Esposito-PRE-2006}.

Fluctuation relations for phase variables other than the dissipation function can be straightforwardly obtained for those phase variables that are odd with respect to time reversal.  These relations are more complex than the dissipation-function FT \cite{Searles-ACS-2000, Evans-AP-2002, Searles-JSP-2007}.  Local fluctuation relations have also be derived \cite{Ayton-JCP-2001,Gallavotti-PA-1999}.

Finally, as the functional form of the dissipation function in the FT depends on the initial ensemble of the system, satisfaction of an FT that is derived assuming a particular distribution function provides evidence that the correct ensemble had been assumed. This has proved useful in the study of glasses \cite{Williams-arXiv-2007}.

Some experimental studies of the FTs have been discussed.  These have allowed confirmation of the relationships in real physical systems.  We see that a challenge now lies in applying the FTs to assist in the development of nanotechnology and to developing our understanding of biological systems. 

\section{Acknowledgements}
We gratefully acknowledge Prof Denis Evans' contribution to this manuscript and his valuable comments.  We thank Dr David Carberry for preparation of the figures, and the Australian Research Council for their support.

\end{document}